\pdfoutput=1
\documentclass[aps,prx,reprint,onecolumn,superscriptaddress]{revtex4-2}
\usepackage[T1]{fontenc}
\usepackage[utf8]{inputenc}
\usepackage{physics} % for DeclareMathOperator
\usepackage{qcircuit}
\usepackage{amsthm} % for theorems
\usepackage{amssymb}
\usepackage{bbold} % to write in blackboard
\usepackage{mathtools} % for DeclarePairedDelimiter
\usepackage{bm}
\usepackage{graphicx} % for figures
\usepackage{subfig}
\usepackage[hidelinks]{hyperref}
\usepackage{natbib}
\usepackage[dvipsnames]{xcolor}

\DeclarePairedDelimiter{\ceil}{\lceil}{\rceil}

\DeclareMathOperator{\polylog}{polylog}
\DeclareMathOperator{\cube}{cube}

\theoremstyle{plain}
\newtheorem{theorem}{Theorem}

\begin{document}

\title{Quantum Simulation of Bound State Scattering}

\author{Matteo Turco}
\affiliation{Physics of Information and Quantum Technologies Group, Centro de Física e Engenharia de Materiais Avançados (CeFEMA), Lisbon, Portugal}
\affiliation{Laboratory of Physics for Materials and Emergent Technologies, Lisbon, Portugal}
\affiliation{Instituto Superior Técnico, Universidade de Lisboa, Lisbon, Portugal}
\affiliation{PQI -- Portuguese Quantum Institute, Lisbon, Portugal}

\author{Gonçalo Quinta}
\affiliation{Instituto de Telecomunicações, Lisbon, Portugal}

\author{João Seixas}
\affiliation{Physics of Information and Quantum Technologies Group, Centro de Física e Engenharia de Materiais Avançados (CeFEMA), Lisbon, Portugal}
\affiliation{Laboratory of Physics for Materials and Emergent Technologies, Lisbon, Portugal}
\affiliation{Departamento de Física, Instituto Superior Técnico, Universidade de Lisboa, Lisbon, Portugal}
\affiliation{PQI -- Portuguese Quantum Institute, Lisbon, Portugal}
https://www.overleaf.com/project/645a3e79dab1c0a6c05f1ca8

\author{Yasser Omar}
\affiliation{Physics of Information and Quantum Technologies Group, Centro de Física e Engenharia de Materiais Avançados (CeFEMA), Lisbon, Portugal}
\affiliation{Laboratory of Physics for Materials and Emergent Technologies, Lisbon, Portugal}
\affiliation{Departamento de Matemática, Instituto Superior Técnico, Universidade de Lisboa, Lisbon, Portugal}
\affiliation{PQI -- Portuguese Quantum Institute, Lisbon, Portugal}

\begin{abstract}
The last few years have seen rapid development of applications
of quantum computation to quantum field theory.
The first algorithms for quantum simulation of scattering
have been proposed in the context of scalar and
fermionic theories, requiring thousands of logical qubits.
These algorithms are not suitable to
simulate scattering of incoming bound states, as the
initial-state preparation relies typically on adiabatically
transforming wavepackets of the free theory
into wavepackets of the interacting theory. In this paper
we present a strategy to excite wavepackets of the
interacting theory directly from the vacuum of the
interacting theory, allowing the preparation of states
of composite particles. This is the first step towards
digital quantum simulation of scattering of bound states.
The approach is based on the Haag-Ruelle scattering theory,
which provides a way to construct creation and annihilation
operators of a theory in a full, nonperturbative framework.
We provide a quantum algorithm requiring
a number of ancillary qubits that is logarithmic in
the size of the wavepackets, and with a success
probability vanishing at most like a polynomial in the lattice
parameters and the energy of the wavepacket.
The gate complexity for a single iteration of the circuit
is equivalent to that of a time evolution for a fixed
time. Furthermore, we propose a complete protocol
for scattering simulation using this algorithm.
We study its efficiency and find improvements with respect to
previous algorithms in the literature.
\end{abstract}

\maketitle

\section{Introduction}
The successes of last few years towards the implementation of
quantum algorithms on real platforms are creating
growing expectation regarding about the opportunities that
quantum computation may open in different fields.
One prominent area that has been particularly fruitful
in providing examples of the potential advantage offered by
quantum computation is high-energy physics~\cite{Bauer-etal_2204}, especially
in what concerns data analysis
\cite{Felser-Trenti-Sestini-Gianelle-Zuliani-Lucchesi-Montangero_2004,Pires-Omar-Seixas_2012,Pires-Bargassa-Seixas-Omar_2101,Magano-Kumar-Kalis-Locans-Glos-Pratapsi-Quinta-Dimitrijevs-Rivoss-Bargassa-Seixas-Ambainis-Omar_2104,Bargassa-Cabos-Choi-Hessel-Cavinato_2106,Delgado-et-al_2203,Belis-Gonzalez-Castillo-Reissel-Vallecorsa-Combarro-Dissertori-Reiter_2104,Schuhmacher-Boggia-Belis-Puljak-Grossi-Pierini-Vallecorsa-Tacchino-Barkoutsos-Tavernelli_2301,Wozniak-Belis-Puljak-Barkoutsos-Dissertori-Grossi-Pierini-Reiter-Tavernelli-Vallecorsa_2301,Crippa-Funcke-Hartung-Heinemann-Jansen-Kropf-Kuhn-Meloni-Spataro-Tuysuz-Yap_2304}
and simulations of lattice quantum field theories
\cite{Jordan-Lee-Preskill_1111,Jordan-Lee-Preskill_1112,Jordan-Lee-Preskill_1404,Preskill_1811,Magnifico-Dalmonte-Facchi-Pascazio-Pepe-Ercolessi_1909,Avkhadiev-Shanahan-Young_2020,Avkhadiev-Shanahan-Young_2209,Klco-Savage-Stryker_1908,Mueller-Tarasov-Venugopalan_1908,Kreshchuk-Kirby-Goldstein-Beauchemin-Love_2002,Haase-Dellantonio-Celi-Paulson-Kan-Jansen-Muschik_2006,Echevarria-Egusquiza-Rico-Schnell_2011,Qian-Basili-Pal-Luecke-Vary_2112,Bauer-Nachman-Freytsis_2102,Atas-Zhang-Lewis-Jahanpour-Haase-Muschik_2102,Cohen-Lamm-Lawrence-Yamauchi_2104,Stetina-Ciavarella-Li-Wiebe_2101,Kan-Nam_2107,Funcke-Hartung-Jansen-Kuhn-Schneider-Stornati-Wang_2110,Bauer-Grabowska_2111,Czajka-Kang-Ma-Zhao_2112,Gonzalez-Cuadra-Zache-Carrasco-Kraus-Zoller_2203,Atas-Haase-Zhang-Wei-Pfaendler-Lewis-Muschik_2207,Farrell-Chernyshev-Powell-Zemlevskiy-Illa-Savage_2207,Farrell-Chernyshev-Powell-Zemlevskiy-Illa-Savage_2209,Davoudi-Mueller-Powers_2208,Davoudi-Mueller-Powers_2212,Ciavarella-Caspar-Illa-Savage_2210,Grabowska-Kane-Nachman-Bauer_2208,Kane-Grabowska-Nachman-Bauer_2211,Kane-Grabowska-Nachman-Bauer_2212,Lamm-Lawrence-Yamauchi_1908,Nachman-Provasoli-deJong-Bauer_1904,Gustafson-Zhu-Dreher-Linke-Meurice_2021,Gustafson-Prestel-Spannowsky-Williams_2207,Mariani-Pradhan-Ercolessi_2301,Funcke-Hartung-Jansen-Kuhn_2302,Chai-Crippa-Jansen-Kuhn-Pascuzzi-Tacchino-Tavernelli_2312}.
Such theories have been used
for many decades as a tool to numerically investigate
several aspects of quantum field theory through the
Euclidean path integral approach, which is a powerful tool in its
range of applicability, but does not cover
the whole class of phenomena that require the study of
real-time evolution.
For such problems the Hamiltonian formulation in
Minkowski spacetime is the natural framework,
but it is hardly approachable with classical
computation.
For this class of phenomena -- and arguably in all the theory of fundamental
interactions -- scattering events are of special
interest since they are essentially the only
means we have to access those regimes of physics where
quantum field theory is necessary.
This work concerns state preparation for digital quantum simulation of scattering.

The prospect of large-scale, fault-tolerant
quantum computers, however far they may be in the future,
has already started to change our approach to lattice field
theory through the seminal
papers of Jordan, Lee, and Preskill
(JLP)~\cite{Jordan-Lee-Preskill_1111,Jordan-Lee-Preskill_1112}.
To exploit the potential advantage offered by quantum
computers, it is better to formulate the theory on a space
lattice with continuous time rather than on a spacetime
lattice, as traditionally done.
In Refs.~\cite{Jordan-Lee-Preskill_1111,Jordan-Lee-Preskill_1112},
an efficient quantum algorithm for simulation
of scattering in the scalar theory $\phi^4$,
requiring thousands of logical qubits,
is provided and analyzed.
The present work aims to
contribute to this long-term perspective by
making the first step towards digital quantum simulation of
scattering events with incoming composite particles.
Preparation of bound states is
an essential task to ultimately perform simulations
of important, real-life collider events, such as
proton-proton scattering at the Large Hadron Collider
(see Ref.~\cite{Rigobello-Notarnicola-Magnifico-Montangero_2105}
for recent developments in the context of tensor
networks
and
Refs.~\cite{Marshall-Pooser-Siopsis-Weedbrook_1503,Surace-Lerose_2011,Karpov-Zhu-Heller-Heyl_2011,Mukherjee-Bastianello-Knolle_2204,Belyansky-Whitsitt-Mueller-Fahimniya-Bennewitz-Davoudi-Gorshkov_2307,Briceno-Edwards-Eaton-Gonzalez-Arciniegas-Pfister-Siopsis_2312}
for recent developments in
the context of analog quantum simulation).

The same problem was treated in
Refs.~\cite{Brennen-Rohde-Sanders-Singh_1412,Bagherimehrab-Sanders-Berry-Brennen-Sanders_2110,Barata-Mueller-Tarasov-Venugopalan_2012},
with use of different, albeit equivalent, formulations such as
the wavelet basis or the multiparticle
decomposition of the Hilbert space. A common feature
of these studies is that they all rely on
excited states of the associated free theory to
prepare wavepackets of the interacting theory. An immediate
consequence is that these approaches can be used only for states
of the interacting theory that can be obtained by smooth interpolation
(typically by an adiabatic transformation)
from states of the free theory. This excludes the
case of bound states.

Having in mind the framework
in Refs.~\cite{Jordan-Lee-Preskill_1111,Jordan-Lee-Preskill_1112},
we provide in the present work a general strategy to
prepare single-particle wavepackets of elementary
or composite particles,
with lower and upper mass gaps, on a quantum
computer. We assume that the preparation of the vacuum state
of the interacting theory is available and that we have access
to an interpolating operator between the vacuum and
the particle we want to create. Once the incoming
wavepackets are created, the time evolution
and measurements steps of the quantum simulation algorithm
proceed as
in Refs.~\cite{Jordan-Lee-Preskill_1111,Jordan-Lee-Preskill_1112}.
The key idea is to use the Haag-Ruelle
scattering theory, which is an alternative
and complementary approach
to the Lehmann-Symanzik-Zimmerman (LSZ) theory.
The Haag-Ruelle formalism is of great
conceptual importance in the context of axiomatic
quantum field theory since it provides
a link between the LSZ framework and the
Wightman axioms (see, for instance, Ref.~\cite{Duncan}
for more details). From an operative point of view,
its success has been rather limited, on one
hand because it is outperformed by the LSZ formalism
in the context of perturbation theory and on the
other hand because real-time evolution, essential
to any scattering theory, is unmanageable with traditional
lattice techniques. Quantum computers may
offer a new boost to the Haag-Ruelle formalism
in terms of operativeness.

The method we propose requires a number of
ancillary qubits that is logarithmic in the
size of the wavepacket. We provide a quantum circuit
with a gate complexity
that is equivalent to that of a time
evolution for a fixed time, and a certain probability of success.
We argue
that this probability does not
vanish faster than a polynomial in the
lattice parameters in the continuum limit.
For definiteness, we work here with a single scalar
field, because it is an illustrative case
and is directly comparable with
what is available in the literature. However, the idea
holds, \textit{mutatis mutandis}, with other theories as well.

State preparation is typically the most difficult
step in digital quantum simulation of scattering,
both technically and in terms of complexity. This
work provides an
innovative approach to the topic and sets the route
to the preparation of composite particles.
Here we consider the problem of preparing the vacuum
only briefly. It is an interesting problem on its own
and has already been addressed in other papers
\cite{Klco-Savage_1912,Li-Macridin-Mrenna-Spentzouris_2210,Paulson-Dellantonio-Haase-Celi-Kan-Jena-Kokail-vanBijnen-Jansen-Zoller-Muschik_2008,Ciavarella-Chernyshev_2112,Farrell-Illa-Ciavarella-Savage_2308,Cohen-Oh_2310}.
For example, one may use the free vacuum preparation
and the adiabatic
transformation in Ref.~\cite{Jordan-Lee-Preskill_1112}, with
the simplification that no backward time evolution is
needed to contain premature wavepacket propagation.
We consider this possibility and study its efficiency, with the
purpose of providing an estimation of the total
scaling of state preparation using our approach.
We also compare our approach with the approach of Jordan, Lee, and
Preskill, and find that our protocol has a comparable or better scaling
than theirs in some cases.
If one is interested in only scattering amplitudes,
a simpler approach not involving quantum simulation
would be the one used in Ref.~\cite{Li-Lai-Wang-Xing_2301}. However that
approach works only for fixed final states and for
processes with a small total number $n$ of ingoing and
outgoing particles, as the complexity scales
exponentially with $n$.

We mentioned previously that we rely on
the existence of lower and upper mass gaps, but in general
we can also have bound states immersed in the continuum of
multiparticle states under the condition that they
are protected by some symmetry. In this case
our strategy is still suitable with some extra caveats.
Fermionic theories do not present extra problems
apart from the ones related to encoding
of anticommuting degrees of freedom on
a qubit system.
Gauge theories, as usual, require special attention,
not only because of the the well-known issues related
to quantum simulation of these theories, but also
because of the formulation of the Haag-Ruelle
theory in the presence of massless particles such as photons.
Nevertheless, we assume that with proper
care these problems can be solved.
For ultrarelativistic particles, the gap closes,
and correspondingly the difficulty of
creating a particle increases with its momentum.
This approach is not suited for the creation of
massless particles, for which there is no mass gap.

The rest of this paper is divided into three main sections and an appendix.
In Sec.~\ref{Haag-Ruelle-theory} we very briefly
introduce the axiomatic approach
to quantum field theory, where the Haag-Ruelle
scattering theory is developed. We list the
Wightman axioms and introduce
basic concepts of the theory. We discuss first
a theory with a single elementary particle, and then proceed to a
theory in one space dimension with both an elementary
particle and a composite one. We end Sec.~\ref{Haag-Ruelle-theory}
with some remarks on the applicability of our work. In Sec.~\ref{wavepacket-creation} we
describe how the Haag-Ruelle scattering theory
can be used for state preparation in digital
quantum simulation. We provide a quantum algorithm
for particle creation from the interacting vacuum,
and we analyze its complexity and probability of success.
In Sec.~\ref{scattering-protocol} we consider a full
protocol for initial-state preparation based on adiabatically
preparing the interacting vacuum and then creating wavepackets from it.
In the Appendix we discuss the truncation of the field operator
in a generic site of the lattice.

\section{The Haag-Ruelle Scattering Theory}
\label{Haag-Ruelle-theory}
The Haag-Ruelle scattering theory is best developed in the
framework of axiomatic quantum field theory.
The axiomatic approach provides a rigorous framework
to construct a quantum field theory. It is
based on a set of axioms formulated by Wightman and
incorporates at the same time the principles of
quantum mechanics and special relativity.
An important feature of this approach is that
free and interacting theories are treated on equal terms.
In particular, interacting theories are not seen as
extensions of free theories obtained by adding
interaction terms to a quadratic Hamiltonian.
The difference between free theories and interacting theories
is mainly of a pragmatical nature, because we
can solve and construct explicitly free theories,
but for most of the interacting theories the same is not true.
For this reason it is quite uncommon to use this approach
operatively.
In the next two sections we review the Haag-Ruelle
formalism, first in the context of a theory with only
an elementary particle, and then in the context of a
theory with bound states.

\subsection{Scalar Field Without Bound States}
In this section we give a brief review
of the axioms following essentially chapter 9
in Ref.~\cite{Duncan}. We consider $d=D-1$ space dimensions, and we consider here
a theory containing a single
scalar field, whose dynamics give rise to a single kind
of particle of mass $m$. Examples of such theories include the free scalar field theory and the
$\phi^4$ theory at weak coupling.
The latter has been shown to satisfy the
axioms for $d=1$ and $d=2$. In the physical case $d=3$,
it is believed to be trivial, \textit{i.e}. equivalent to
a free theory, while for $d\ge4$, triviality has been proven
\cite{Glimm-Jaffe_68,Glimm-Jaffe_70-1,Glimm-Jaffe_70-2,Glimm-Jaffe_72,Glimm-Jaffe_73,Glimm-Jaffe-Spencer_74}
(see also the discussion at the beginning
of Sec. 2.1 in Ref.~\cite{Jordan-Lee-Preskill_1112}).
The free theory is known to satisfy the axioms.
The framework can be adapted
straightforwardly to the case of a theory with fermionic
fields. All that is required is to change
commutation relations into anticommutation relations.

It is convenient to divide the whole
set of axioms into a few families according
to their content.
The first one concerns the space of states and the
spectral properties of the theory:
\begin{itemize}
\item[1.] \textbf{Axiom Ia} The state space $\mathcal{H}$
of the system is a separable Hilbert space. It carries
a unitary representation $U(\Lambda,x)$
($\Lambda$ is an element of the homogeneous Lorentz group
and
$x$ is a spacetime coordinate vector) of the proper inhomogeneous
Lorentz group (i.e., the Poincaré group).
Thus, for all $\ket{\alpha}\in\mathcal{H}$,
$\ket{\alpha}\to U(\Lambda,x)\ket{\alpha}$, with the
$U(\Lambda,x)$ satisfying the
Poincaré algebra
\begin{equation}
U(\Lambda_1,x_1)U(\Lambda_2,x_2)=
U(\Lambda_1\Lambda_2,x_1+\Lambda_1x_2).
\end{equation}
\item[2.] \textbf{Axiom Ib} The infinitesimal generators
$P_{\mu}$ of the translation subgroup $T(x)=U(\openone,x)$
of the Poincaré group have a spectrum $p_{\mu}$ restricted
to the forward light cone, $p_0\ge0,\,p^2\ge0$.
\item[3.] \textbf{Axiom Ic}
There is a unique state $\ket{\Omega}$,
the vacuum, with the isolated eigenvalue $p_{\mu}=0$
of $P_{\mu}$.
\item[4.] \textbf{Axiom Id} \label{axiomId}
The theory has a mass gap:
the squared-mass operator
\begin{equation}
P^2=P_{\mu}P^{\mu}
\end{equation}
has an isolated eigenvalue $m^2>0$, and the spectrum
of $P^2$ is empty between $0$ and $m^2$.
The subspace of $\mathcal{H}$ corresponding to the
eigenvalue $m^2$ carries an
irreducible spin-0 representation of the homogeneous Lorentz group.
These are the single-particle states
of the theory. The remaining spectrum of $P^2$
is continuous, and it begins at $(2m)^2$.
\end{itemize}
Clearly the specific form of the generators $P_{\mu}$
depends on the theory at hand and the
generator of translations in time is the
Hamiltonian, $P_0=H$. One-particle states can be labeled
according to their momentum and can be written as
wavepackets in momentum space
\begin{equation}
\ket{\alpha}_1=\int\!d^dk\,
\tilde{\psi}(\mathbf{k})\ket{\mathbf{k}}
\end{equation}
with some wave function $\tilde{\psi}$.

The next family of axioms concerns the operator
content of the Hilbert space and establishes what kind
of fields appear in the theory:
\begin{enumerate}
\item[5.] \textbf{Axiom IIa} \label{axiomIIa}
An operator-valued (tempered)
distribution $\hat{\phi}(x)$ exists such that
for any Schwartz test function $f(x)$ the smeared field
\begin{equation}
\phi_f=\int\!d^Dx\,f(x)\hat{\phi}(x)
\end{equation}
is an unbounded operator defined on a dense subset
$D\subset\mathcal{H}$. Moreover, $\phi_fD\subset D$,
allowing the definition of arbitrary (finite) products of
smeared fields.
\end{enumerate}
We recall that a Schwartz function $f(x)$
is infinitely differentiable  (i.e. is $C^{\infty}$)
and, together with all its derivatives, falls
faster than any power of $x$ as $x$ goes to infinity.
\begin{enumerate}
\item[6.] \label{axiomIIb}
\textbf{Axiom IIb} Under the unitary
representation of the Poincaré group
$U(\Lambda,x)$ introduced in \textbf{Axiom Ia},
the smeared fields transform as
\begin{equation}
U(\Lambda,x)\phi_fU^{\dagger}(\Lambda,x)=
\int\!d^Dy\,f\left[\Lambda^{-1}(y-x)\right]\hat{\phi}(y).
\end{equation}
\item[7.] \textbf{Axiom IIc} Let $f_1$ and $f_2$ be Schwartz
functions of compact support: thus, if $f_1$
vanishes outside a compact spacetime region $v_1$
and $f_2$ vanishes outside
the compact region $v_2$, and if $x_1-x_2$
is space-like for all $x_1\in v_1$, $x_2\in v_2$, then
\begin{equation}
\big[\phi_{f_1},\phi_{f_2}\big]=0.
\end{equation}
\item[8.] \textbf{Axiom IId} The set of states obtained
by applying arbitrary polynomials in the
smeared fields $\phi_f$ (with all possible
Schwartz functions $f$) to the vacuum state $\ket{\Omega}$
is dense in the Hilbert space $\mathcal{H}$.
\end{enumerate}
In \hyperref[axiomIIa]{\textbf{Axiom IIa}}
we introduce an important
difference with respect to canonical quantization, namely smeared
operators. In a free theory we can identify
the operator-valued
distribution $\hat{\phi}(x)$ with
the familiar field operator. This is not
a well-defined operator because the state
$\hat{\phi}(x)\ket{\Omega}$ has an infinite norm. To avoid this
problem, it is necessary to introduce smearing
and treat $\hat{\phi}(x)$ as a
distribution. In this way the state $\phi_f\ket{\Omega}$
has a finite norm for any Schwartz function $f$ and
$\phi_f$ is a well-defined (unbounded) operator.
Then, by \hyperref[axiomIIb]{\textbf{Axiom IIb}},
we can define
\begin{equation}
\phi_f(x)=
e^{iP\cdot x}\phi_fe^{-iP\cdot x}=
\int\!d^Dy\,f(y-x)\hat{\phi}(y).
\label{almost-local-operator}
\end{equation}

There are two more axioms of great importance to
develop a satisfactory scattering theory:
\begin{enumerate}
\item[9.] \textbf{Axiom IIIa} \label{axiomIIIa}
For some one-particle state
\begin{equation}
\ket{\alpha}_1=\int\!d^dk\,\tilde{\psi}(\mathbf{k})\ket{\mathbf{k}}
\end{equation}
with discrete eigenvalue $m^2$ of the squared-mass operator,
the smeared
field $\phi_f(x)$ has a nonvanishing matrix
element from this single-particle state to
the vacuum,
\begin{equation}
\braket{\Omega}{\phi_f(x)|\alpha}_1\ne0.
\end{equation}
\item[10.] \textbf{Axiom IIIb} \label{axiomIIIb}
(asymptotic completeness)
The Hilbert space $\mathcal{H}_{\text{in}}$
(or $\mathcal{H}_{\text{out}}$)
corresponding to multiparticle states of far-separated,
freely moving stable
particles in the far past (or far future)
are unitarily equivalent, and may be
identified with the full Hilbert space $\mathcal{H}$
of the system.
\end{enumerate}
It should be noted that
\hyperref[axiomIIIb]{\textbf{Axiom IIIb}} plays a crucial
role in the derivation of the LSZ reduction formula,
but here it is somewhat superfluous.

\begin{figure}
\includegraphics[width=0.47\textwidth]{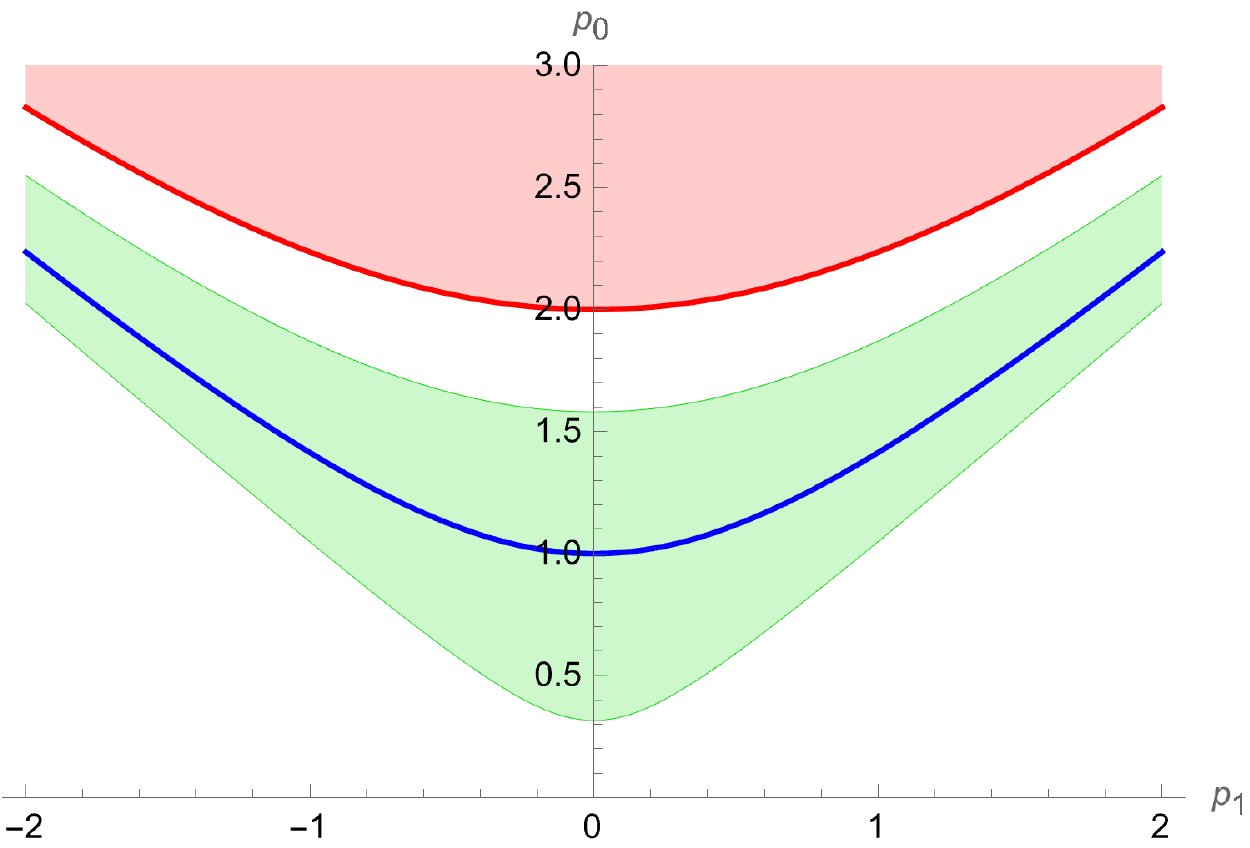}
\caption{\label{P-spectrum}Structure
of the spectrum of $P_{\mu}$
with one space dimension. The blue line represents
the one-particle mass hyperboloid, the red region
is the multiparticle continuum, and the green
region is the region defined by $am^2<p^2<bm^2$.}
\end{figure}

The joint set of eigenvalues of $P_{\mu}$, labeled by
$p_{\mu}$, is composed of three disconnected
subsets (see Figure~\ref{P-spectrum}).
There is the vacuum subset, containing only
the origin $p_{\mu}=0$, the
one-particle mass hyperboloid, containing all
the $p_{\mu}$ points such that $p^2=m^2$, and the multiparticle continuum with
all the points such that $p^2\ge 4m^2$. In the two-particle subspace, for instance,
the squared-mass operator gives
$(p_1+p_2)^2=2m^2+2p_1\cdot p_2$, with
$p_1\cdot p_2\ge m^2$.

With this in mind we define an operator $\phi_1(x)$
exactly as in Eq.~\eqref{almost-local-operator} but with
a smearing function $f_1(x)$ chosen as the Fourier
transform of a function $\tilde{f}_1(p)$
with support in the region
\begin{equation}
am^2<p^2<bm^2,\qquad0<a<1<b<4,
\end{equation}
sandwiching the one-particle
mass hyperboloid. This guarantees that the state
$\phi_1(x)\ket{\Omega}$ is a one-particle (and one-particle
only) state, by \hyperref[axiomIIIa]{\textbf{Axiom IIIa}}.

Next consider a positive-energy solution of the
Klein-Gordon equation,
\begin{equation} %onecolumn
g(\tau,\mathbf{y})=\int\!
\frac{d^dp}{2E(\mathbf{p})}
\tilde{g}(\mathbf{p})
e^{i(\mathbf{p}\cdot\mathbf{y}-E(\mathbf{p})\tau)},\qquad
E(\mathbf{p})=\sqrt{m^2+|\mathbf{p}|^2},
\end{equation}
%\begin{align}
%g(\tau,\mathbf{y})&=\int\!
%\frac{d^dp}{2E(\mathbf{p})}
%\tilde{g}(\mathbf{p})
%e^{i(\mathbf{p}\cdot\mathbf{y}-E(\mathbf{p})\tau)},\\
%E(\mathbf{p})&=\sqrt{m^2+|\mathbf{p}|^2},
%\end{align} %twocolumn
and we define the operator
\begin{equation}
\phi_{1,g}(\tau)=-i\int\!d^dy\,
\bigg[
g(\tau,\mathbf{y})
\overleftrightarrow{\frac{\partial}{\partial\tau}}
\phi_1(\tau,\mathbf{y})
\bigg],
\label{one-particle-creation}
\end{equation}
where the double derivative is defined as in
\begin{equation}
A(\tau)
\overleftrightarrow{\frac{\partial}{\partial\tau}}B(\tau)=
A(\tau)\dot{B}(\tau)-\dot{A}(\tau)B(\tau).
\end{equation}
The state $\phi_{1,g}(\tau)\ket{\Omega}$ can be shown
to be independent of $\tau$,
\begin{equation}
\frac{d}{d\tau}\phi_{1,g}(\tau)\ket{\Omega}=0.
\label{independence-on-tau}
\end{equation}
This is no longer true in general
if we consider multiple applications of such operators
at the same time $\tau$,
$\phi_{1,g_1}(\tau)\cdots\phi_{1,g_2}(\tau)\ket{\Omega}$,
but in this case we can rely on the following theorem
(see chapter 9 in Ref.~\cite{Duncan} for more details).
\begin{theorem}[Haag Asymptotic]
The time-dependent state vector
\begin{equation}
\ket{\Psi,\tau}=
\phi_{1,g_1}(\tau)\cdots\phi_{1,g_n}(\tau)\ket{\Omega}
\label{Haag-state}
\end{equation}
converges strongly in the limit $\tau\to-\infty$ to
the $n$-particle in-state
\begin{equation} %onecolumn
\ket{\Psi}_{\text{in}}=
\ket{g_1,g_2,\dots,g_n}_{\text{in}}=
\int\!d^dp_1\dots d^dp_n\,
\tilde{\psi}_{1,g_1}(\mathbf{p}_1)\cdots
\tilde{\psi}_{1,g_n}(\mathbf{p}_n)
\ket{\mathbf{p}_1\dots\mathbf{p}_n}_{\text{in}},
\end{equation}
%\begin{multline}
%\ket{\Psi}_{\text{in}}=
%\ket{g_1,g_2,\dots,g_n}_{\text{in}}=\\
%\int\!d^dp_1\dots d^dp_n\,
%\tilde{\psi}_{1,g_1}(\mathbf{p}_1)\cdots
%\tilde{\psi}_{1,g_n}(\mathbf{p}_n)
%\ket{\mathbf{p}_1\dots\mathbf{p}_n}_{\text{in}},
%\end{multline} %twocolumn
with momentum wave functions
\begin{equation}
\tilde{\psi}_{1,g_i}(\mathbf{p}_i)=
(2\pi)^{d/2}
\frac{\tilde{g}(\mathbf{p}_i)
\tilde{f}_1(\mathbf{p}_i)}{\sqrt{2E(p_i)}}.
\end{equation}
\label{Haag asymptotic theorem}
\end{theorem}
The convergence rate of the limit
$\ket{\Psi,\tau}\to\ket{\Psi}_{\text{in}}$ is
$|\tau|^{-d/2}$ in the general case, which includes
the case where some or all of the wavepackets
$g_1,\dots,g_n$ overlap with each other and
come from the same direction.
In this case, the convergence
is guaranteed by spreading of the wavepackets to
the point where they are so broad that they cease to interact.
In practice we consider only wavepackets coming
from different directions, and in this case the fast
decrease of the Schwartz functions ensures that
the convergence rate is faster than any inverse
power of $|\tau|$.

The state $\ket{\Psi}_{\text{in}}$ is a Heisenberg state,
which implies that it is not to be visualized in general as
a state made of $n$ (spatially) well-separated wavepackets.
Its form depends on the reference time
at which the Heisenberg
state and the Schrödinger state coincide.
As the reference time,
we can choose a moment well before the collision between
the wavepackets occurs, in which case we indeed
have well-separated wavepackets, or a moment during
the collision or later, in which
case we can expect to have a complicated state
more or less spread in space. The strong convergence
of the theorem is to be taken at the same reference time,
whether in the far past or not, for both
$\ket{\Psi,\tau}$ and $\ket{\Psi}_{\text{in}}$. This reference time should not be confused with
the parameter $\tau$ appearing in the theorem.
But since we will work in the Schrödinger picture this
subtlety is not relevant to us.

Let us see more explicitly what we have stated so far,
starting from the state
\begin{equation}
\phi_{1,g}(\tau)\ket{\Omega}=
\int\!d^Dx\,\psi(x;\tau)
\hat{\phi}(x)\ket{\Omega},
\label{one-particle-state}
\end{equation}
where
\begin{equation} %onecolumn
\psi(x;\tau)=(2\pi)^d\int\!d^Dp\,
\tilde{g}(\mathbf{p})\tilde{f}_1(p)
\frac{p_0+E(\mathbf{p})}{2E(\mathbf{p})}
e^{-i\tau[E(\mathbf{p})-p_0]}
e^{-ip\cdot x}
\label{wavefunction}
\end{equation}
%\begin{multline}
%\psi(x;\tau)=\\
%(2\pi)^d\int\!d^Dp\,
%\tilde{g}(\mathbf{p})\tilde{f}_1(p)
%\frac{p_0+E(\mathbf{p})}{2E(\mathbf{p})}
%e^{-i\tau[E(\mathbf{p})-p_0]}
%e^{-ip\cdot x}
%\label{wavefunction}
%\end{multline} %twocolumn
is obtained from~\eqref{one-particle-creation}
after some simple manipulations.
For our purposes it is convenient to move to
the Schrödinger picture by writing $x=(t,\mathbf{x})$
and plugging
\begin{equation}
\hat{\phi}(x)=
e^{itH}\hat{\phi}(0,\mathbf{x})e^{-iHt}=
e^{itH}\hat{\phi}(\mathbf{x})e^{-iHt}
\end{equation}
into~\eqref{one-particle-state}. Furthermore,
shifting the integration variable in~\eqref{one-particle-state}
by $t\to t+\tau$, we get
\begin{gather}
\phi_{1,g}(\tau)\ket{\Omega}=
e^{iH\tau}a_{\psi}^{\dagger}(\tau)\ket{\Omega},
\label{Haag-Ruelle-creation}\\
a_{\psi}^{\dagger}(\tau)=\int\!d^Dx\,\psi(t+\tau,\mathbf{x};\tau)
e^{iHt}\hat{\phi}(\mathbf{x})e^{-iHt}.
\end{gather}
Let us have a closer look at $\psi(t+\tau,\mathbf{x};\tau)$.
We can rewrite it as
\begin{gather}
\psi(t+\tau,\mathbf{x};\tau)=(2\pi)^d\int\!d^dp\,
\tilde{g}(\mathbf{p})
\tilde{f}'_1(\mathbf{p};t)
e^{i[\mathbf{x}\cdot\mathbf{p}-\tau E(\mathbf{p})]},
\label{smeared-Klein-Gordon}\\
\tilde{f}'_1(\mathbf{p};t)=\int_{-\infty}^{+\infty}
\!dp_0\,\tilde{f}_1(p)
\frac{p_0+E(\mathbf{p})}{2E(\mathbf{p})}e^{-itp_0}.
\end{gather}
If, without loss of generality, we assume
$f_1(t,\mathbf{x})$ peaked at around $t=0$, so is
$\tilde{f}'_1(\mathbf{p};t)$. Roughly speaking,
the effect of $\tilde{f}'_1(\mathbf{p};t)$ is to spread
$\psi(t+\tau,\mathbf{x};\tau)$ without changing its position.
Therefore, by~\eqref{smeared-Klein-Gordon} we
see that $\psi(t+\tau,\mathbf{x};\tau)$ is
essentially a solution
of the Klein-Gordon equation moving through
space with time $\tau$. When $\tau<0$,
$a_{\psi}^{\dagger}(\tau)$ creates a wavepacket at
some point along the past trajectory of
$\psi(t+\tau,\mathbf{x};\tau)$. Then,
the time evolution operator
in~\eqref{Haag-Ruelle-creation} moves it
to where $\psi(t+\tau,\mathbf{x};\tau)$ lies at $\tau=0$.
This accounts for~\eqref{independence-on-tau}.

Similarly, in the case of two incoming particles we can
write
\begin{equation}
\phi_{1,g_1}(\tau)\phi_{1,g_2}(\tau)\ket{\Omega}=
e^{iH\tau}a_{\psi_1}^{\dagger}(\tau)a_{\psi_2}^{\dagger}(\tau)
\ket{\Omega}.
\label{2-particles-creation}
\end{equation}
We can choose $g_1(\tau,\mathbf{y}_1)$
and $g_2(\tau,\mathbf{y}_2)$ such that
their wavepackets $\psi_1$ and $\psi_2$ are always
well separated from each other for $\tau\le0$, and
are on a collision course for some $\tau>0$.
Then, from~\eqref{2-particles-creation},
the action of the operators
$\phi_{1,g_1}(\tau)\phi_{1,g_2}(\tau)$ on the vacuum is clear:
as $\tau\to-\infty$, the two wave functions
$\psi_1(t_1+\tau,\mathbf{x}_1;\tau)$ and
$\psi_2(t_2+\tau,\mathbf{x}_2;\tau)$ are sent far from
each other where the two creation operators
$a_{\psi_1}^{\dagger}(\tau)$ and
$a_{\psi_2}^{\dagger}(\tau)$ act undisturbed by
each other. Then, the time evolution
operator $e^{iH\tau}$ evolves the system forward
making the two wavepackets approach and interact
with each other.
Let us call $r$ the distance between the regions
where $\psi_1(t_1+\tau,\mathbf{x}_1;\tau)$ and
$\psi_2(t_2+\tau,\mathbf{x}_2;\tau)$
are concentrated at $\tau=0$.
Provided that interactions between particles
are short-ranged in the theory, as is the case
for gapped theories, we can ignore the
interactions occurring between the two wavepackets
during the evolution from
$\tau=-\infty$ to $\tau=0$, with error
vanishing faster than any inverse power of $r$.
Thus, at the cost of slightly increasing $r$,
we can take
$\tau=0$ in~\eqref{2-particles-creation}
with excellent precision.

\subsection{Scalar Field with Bound States}
The theory with a single scalar field with interactions
$\lambda(\hat{\phi}^6-\hat{\phi}^4)$,
in a single space dimension,
is a simple example of a quantum field theory
displaying bound states. This theory is known to satisfy
the Wightman axioms \cite{Glimm-Jaffe-Spencer_74}, and
at weak coupling, it has a single composite particle below the
two-particle threshold $2m$. Its mass $m_{\text{b}}$
can be computed perturbatively
\cite{Dimock-Eckmann_76,Spencer-Zirilli_76,Dimock-Eckmann_77}
as
\begin{equation}
m_{\text{b}}=2m\bigg[
1-\frac{9}{8}\bigg(\frac{\lambda}{m^2}\bigg)^2+O(\lambda^3)
\bigg].
\end{equation}
The bare mass $m_0(\lambda)$ can be set such that
$m=m(m_0(\lambda),\lambda)$ is fixed.
We notice at this
point that \hyperref[axiomId]{\textbf{Axiom Id}}
should be modified in an obvious way to accommodate
the composite particle with mass $m_{\text{b}}$.
This is a well-controlled model that
is perfect to see in a simple and explicit way how
to treat bound states in the Haag-Ruelle formalism.
The last ingredient we need to know is that
the field $:\hat{\phi}^2:(x)$ interpolates between
the vacuum and one-particle states of the composite
particle \cite{Dimock-Eckmann_77}. The Wick ordering
$:\cdot:$ is necessary in the continuum to avoid
divergences of $\hat{\phi}^2(x)$, but on the lattice
it amounts to a constant, finite shift. This is
completely irrelevant to our purposes as can
be easily seen. Let $\hat{\Phi}(x)$ denote either
$\hat{\phi}(x)$ or $:\hat{\phi}^2:(x)$. Then,
if we shift it
by a constant $A$ and smear it with the function
$\psi$ in~\eqref{wavefunction}, we obtain
\begin{equation}
\int\!d^Dx\,\hat{\Phi}(x)\psi(x;\tau)\ket{\Omega}+
A\int\!d^Dx\,\psi(x;\tau)\ket{\Omega}.
\end{equation}
The second term is proportional to
$\tilde{f}_1(0)$, which is zero by construction.

\begin{figure}
\includegraphics[width=0.47\textwidth]{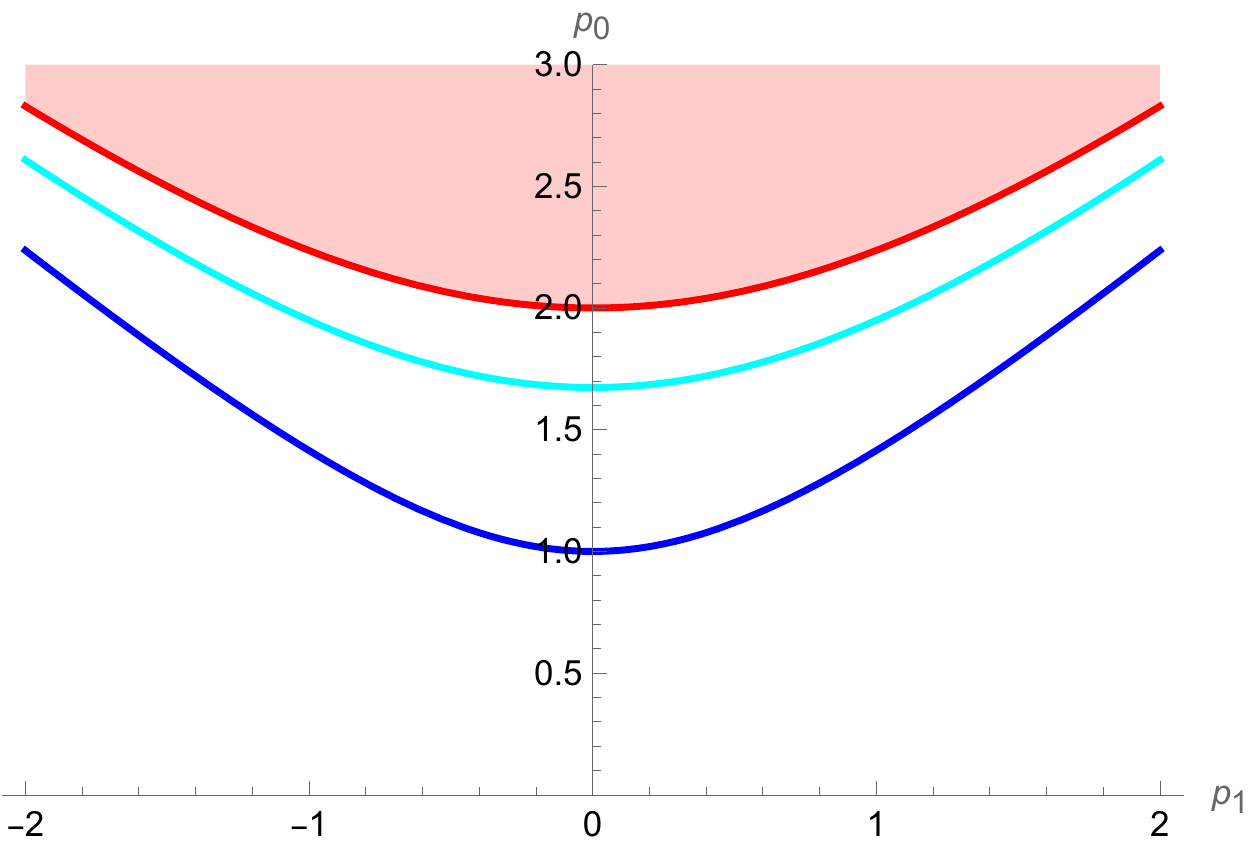}
\caption{\label{P-spectrum_bound-state}Structure
of the spectrum of $P_{\mu}$ in the theory
$\lambda(\hat{\phi}^6-\hat{\phi}^4)$.
The blue line represents
the mass hyperboloid of elementary particles,
the cyan line represents the mass hyperboloid of
two-particle bound states, and the red region
is the multiparticle continuum.}
\end{figure}

The joint spectrum of $P_0=H$ and $P_1$ is depicted
in figure~\ref{P-spectrum_bound-state}.
Starting from this picture, we find that the Haag-Ruelle theory
discussed in the previous section holds equally
well to obtain wavepackets of the elementary particle
with mass $m$, or wavepackets of the composite
particle with mass $m_{\text{b}}$.
If we want to have an elementary particle,
we use the interpolating field $\hat{\phi}(x)$,
a smearing function $f_1(x)$, with Fourier transform
$\tilde{f}_1(p)$ sandwiching the hyperboloid
of mass $m$, and a solution $g(x)$ of the Klein-Gordon
equation to build the operator
$\phi_{1,g}(\tau)$ of
equation~\eqref{one-particle-creation} as before.
If, on the other hand,
we want to obtain a composite particle,
we use the interpolating field $:\hat{\phi}^2:(x)$,
a smearing function $f_{\text{b}}(x)$, with Fourier transform
$\tilde{f}_{\text{b}}(p)$ sandwiching the hyperboloid
of mass $m_{\text{b}}$, and a solution
$h(x)$ of the Klein-Gordon equation to obtain
the operator
\begin{equation}
\phi_{\text{b},h}(\tau)=-i\int_{-\infty}^{+\infty}\!d\mathbf{y}\,
\bigg[
h(\tau,\mathbf{y})
\overleftrightarrow{\frac{\partial}{\partial\tau}}
\phi_{\text{b}}(\tau,\mathbf{y})
\bigg],
\label{bound-state-creation}
\end{equation}
where
\begin{equation}
\phi_{\text{b}}(x)=
\int\!d^2y\,f_{\text{b}}(y-x):\hat{\phi}^2:(y).
\end{equation}
Then everything proceeds as before. A useful
remark is that the field $\hat{\phi}(x)$ does
not interpolate between the vacuum and one-particle
states of the composite kind, and vice versa,
$:\hat{\phi}^2:(x)$ does not interpolate
between the vacuum and one-particle states of
the elementary kind. In the formulae, if $\ket{\alpha_1}$
is an elementary one-particle state and
$\ket{\alpha_{\text{b}}}$ is a composite one-particle state,
we have
\begin{equation}
\bra{\alpha_1}:\hat{\phi}^2:(x)\ket{\Omega}=0=
\bra{\alpha_{\text{b}}}\hat{\phi}(x)\ket{\Omega}.
\label{no-interpolations}
\end{equation}
The first equality
holds because the one-particle sector
of the elementary kind is spanned by vectors of the
form $\phi_1(x)\ket{\Omega}$. Then
the matrix element
$\bra{\alpha_1}:\hat{\phi}^2:(x)\ket{\Omega}$
is an integral containing $\bra{\Omega}\hat{\phi}\,\hat{\phi}\,
\hat{\phi}\ket{\Omega}$, which is zero because the
Hamiltonian contains only even powers of the
field $\hat{\phi}$. A similar argument
leads to $\bra{\alpha_{\text{b}}}\hat{\phi}(x)\ket{\Omega}=0$.
As a consequence of the equalities in Eq.~\eqref{no-interpolations},
$\tilde{f}_1$ and $\tilde{f}_{\text{b}}$ are allowed to have
supports intersecting both hyperboloids of mass $m$ and
$m_{\text{b}}$.

We conclude this section with some remarks on
the validity of the theory just described.
In general, we can have bound states
whose mass falls above the two-particle threshold
of lighter particles, on the condition that
they are protected by some symmetry (internal quantum numbers).
In this case
the symmetry selects a sector of the Hilbert space,
and when we restrict the mass-squared operator
to this sector, the mass of the composite particle
appears as a discrete point in the spectrum of $P^2$ again.

The Haag asymptotic theorem critically depends
on two assumptions:
\begin{enumerate}
\item There exists a lower and an upper mass
gap for the particle we want to create in such a way
that it is possible to sandwich the mass hyperboloid
corresponding to such a particle.
\item We have access to an operator interpolating
between the vacuum and one-particle states of the
particle we want to create.
\end{enumerate}
Clearly, studying these two conditions
strongly depends on the theory under consideration
and can be very difficult, but we can use standard
techniques of lattice quantum field theory to
study these properties model by model, or the Bethe-Salpeter
equation as done
in Refs.~\cite{Dimock-Eckmann_76,Dimock-Eckmann_77,Spencer-Zirilli_76}.
Smearing a field operator in time can be avoided
if we have at our disposal an operator that
does not couple the vacuum to multiparticle
states (as happens for free theories).
For a bound state whose mass is immersed in the
continuum of other particles, we also need to ensure
that the interpolating operator couples only
to the sector where the bound state lives,
in a way similar to the reasoning leading
to~\eqref{no-interpolations}. As an example, consider
a theory where
we have a composite particle with spin $0$ and mass
$m$ (a pion), and another composite particle
with spin $1/2$ and mass $M>2m$ (a proton). These
two particles are made of the same underlying
elementary fields, but the difference in spin should
make it easy to build interpolating fields
from the vacuum to each of the two particles without crossing,
even though the heavier particle has mass in the
continuum of the lighter one.

\section{State Preparation Using the Haag-Ruelle Theory}
\label{wavepacket-creation}

\begin{figure*}
\subfloat[\label{circuit1}\textbf{Circuit 1:}
High-level overview of the circuit implementing
$\bar{a}_{\psi}^{\dagger}$.
The operators $\Phi(t_i)$ are described in the circuit
below.]{\Qcircuit @C=2.8mm @R=5mm {
\lstick{\text{anc.}} &{/}\qw &\gate{V} &\multigate{4}{\Phi(t_1)} &\qw &\multigate{4}{\Phi(t_2)} &\qw &\dots & &\qw &\multigate{4}{\Phi(t_N)} &\gate{V'} &\meter\\
\lstick{\mathbf{x}_1} &{/}\qw &\multigate{5}{e^{-iHt_1}} &\ghost{\Phi(t_1)} &\multigate{5}{e^{iH(t_1-t_2)}} &\ghost{\Phi(t_2)} &\qw &\dots & &\multigate{5}{e^{iH(t_{N-1}-t_N)}} &\ghost{\Phi(t_N)} &\multigate{5}{e^{iHt_N}} &\qw\\
\lstick{\mathbf{x}_2} &{/}\qw &\ghost{e^{-iHt_1}} &\ghost{\Phi(t_1)} &\ghost{e^{iH(t_1-t_2)}} &\ghost{\Phi(t_2)} &\qw &\dots & &\ghost{e^{iH(t_{N-1}-t_N)}} &\ghost{\Phi(t_N)} &\ghost{e^{iHt_N}}&\qw\\
\lstick{\vdots}\\
\lstick{\mathbf{x}_S} &{/}\qw &\ghost{e^{-iHt_1}} &\ghost{\Phi(t_1)} &\ghost{e^{iH(t_1-t_2)}} &\ghost{\Phi(t_2)} &\qw &\dots & &\ghost{e^{iH(t_{N-1}-t_N)}} &\ghost{\Phi(t_N)} &\ghost{e^{iHt_N}} &\qw\\
\\
\lstick{\Gamma} &{/}\qw &\ghost{e^{-iHt_1}} &\qw &\ghost{e^{iH(t_1-t_2)}} &\qw &\qw &\dots & &\ghost{e^{iH(t_{N-1}-t_N)}} &\qw &\ghost{e^{iHt_N}} &\qw
}}\\
\subfloat[\label{circuit2}\textbf{Circuit 2:}
Overview of the operator implementing $\hat{\phi}_{\psi}(t_i)$.
The symbol connecting $\hat{\phi}$ in a
squared box to the rounded
box containing $t_i,\mathbf{x}_j$ represents
the operator $\hat{\phi}(\mathbf{x})$ controlled on
the state
$\ket{t_i,\mathbf{x}_j}$.]{\Qcircuit @C=0.7cm @R=0.5cm {
\lstick{\text{anc.}} &{/}\qw &\qw &\measure{t_i,\mathbf{x}_1}	&\measure{t_i,\mathbf{x}_2}	&\qw &{\dots} & &\measure{t_i,\mathbf{x}_S} &\qw &\qw\\
\lstick{\mathbf{x}_1} &{/}\qw&\multigate{5}{e^{-iHt_i}}&\sgate{\hat{\phi}}{-1}&\qw&\qw&{\dots}& &\qw&\multigate{5}{e^{iHt_i}}&\qw\\
\lstick{\mathbf{x}_2} &{/}\qw &\ghost{e^{-iHt_i}}& \qw &\sgate{\hat{\phi}}{-2}&\qw&{\dots}& &\qw&\ghost{e^{iHt_i}}&\qw\\
\lstick{\vdots} & & & & & & \ddots & & & &\\
\lstick{\mathbf{x}_S} &{/}\qw&\ghost{e^{-iHt_i}}& \qw & \qw & \qw & {\dots} & & \sgate{\hat{\phi}}{-4}&\ghost{e^{iHt_i}}&\qw\\
& & & & {\Phi(t_i)}\\
\lstick{\Gamma} &{/}\qw&\ghost{e^{-iHt_i}}&\qw&\qw&\qw&\dots& & \qw&\ghost{e^{iHt_i}}&\qw
\gategroup{1}{4}{5}{9}{3mm}{--}
}}
\caption{\label{quantum-circuit}Description
of the circuit implementing
$\bar{a}_{\psi}^{\dagger}$. The slash at the beginning of each line
means that the line represents a register of qubits:
$\text{"anc."}$ is the ancillary register of
$N_a$ qubits,
$\mathbf{x}_i$ is the register of $k$ qubits
dedicated to the site $\mathbf{x}_i$, and
$\Gamma$ is the set of qubits dedicated
to the rest of the lattice.}
\end{figure*}
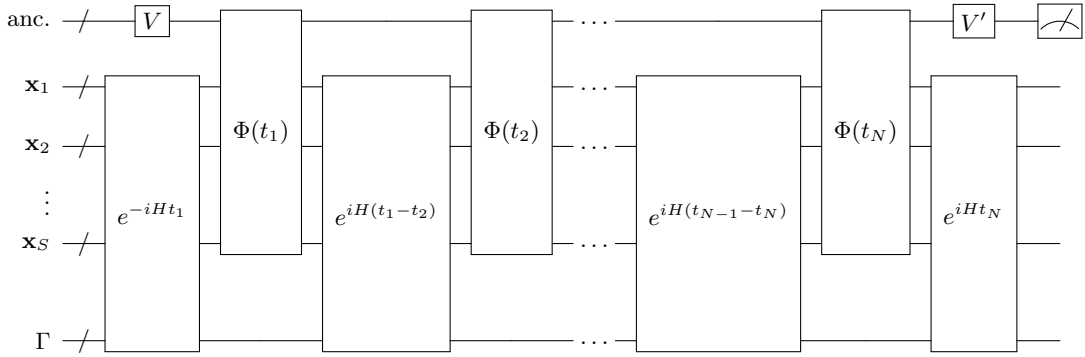
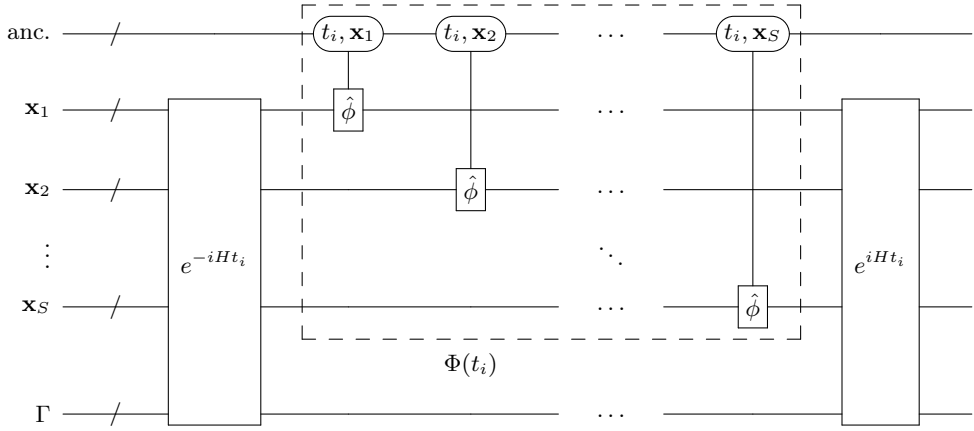

In the following we assume that all of this holds
as an approximation on the lattice.
The Haag-Ruelle scattering theory was
developed for spin systems
in Ref.~\cite{Bachmann-Dybalski-Naaijkens_1412}
and for Euclidean lattice field theory
in Ref.~\cite{Barata-Fredenhagen_91}.
Detailed studies of the effects of latticization
on applications for quantum
simulation will be the subject of future work.

From now on we take $\tau=0$, $\psi(t,\mathbf{x})=\psi(t,\mathbf{x};0)$,
as given in~\eqref{wavefunction}, and
\begin{equation}
a_{\psi}^{\dagger}=\sum_{\mathbf{x}}a^d
\int_{-\infty}^{+\infty}\!dt\,
\psi(t,\mathbf{x})
e^{iHt}\hat{\phi}(\mathbf{x})e^{-iHt}.
\end{equation}
We want to prepare the state
$a_{\psi_1}^{\dagger}a_{\psi_2}^{\dagger}\ket{\Omega}$,
with $\psi_1$ and $\psi_2$ well separated. Because of
this, we can focus on one wavepacket at a time,
so in the following we will see
how to implement the operator $a_{\psi}^{\dagger}$
on a quantum computer.
In~\eqref{wavefunction}, we choose $\tilde{g}$ peaked at around
$\bar{\mathbf{p}}$ with support of size $\delta_p$ and
$\tilde{f}_1(p)$ peaked at around
$(\bar{E},\bar{\mathbf{p}})$ with support of size
$\delta_E$ in the $p_0$ direction and $\delta_p$
in the other directions, with
$\bar{E}=E(\bar{\mathbf{p}})$. Also,
we can take $\delta_E=O(m)$.
First we
truncate the integration over $t$
and the summation over $\mathbf{x}$
around the spacetime region where $\psi$ is significantly
different from zero, which,
since $\psi$ is a Schwartz function, introduces an
error vanishing faster than any power as
the hypervolume of the region is increased.
We label the space points in this region
by $\mathbf{x}_1,\dots,\mathbf{x}_S$
and we approximate
the integral with a sum over time points
$t_1,\dots,t_N$ with spacing $\delta_t$.
By the uncertainty principle, the linear size of $\psi$ over
space is proportional to $1/\delta_p$,
and $t_N-t_1$ is proportional to
$1/\delta_E=O(1/m)$. Thus, as an approximation of
$a_{\psi}^{\dagger}$, we have
\begin{equation} %onecolumn
\bar{a}_{\psi}^{\dagger}=
\sum_{i=1}^N\sum_{j=1}^Sa^d\delta_t
\psi(t_i,\mathbf{x}_j)
e^{iHt_i}\hat{\phi}(\mathbf{x}_j)e^{-iHt_i}=
\sum_{i=1}^N\hat{\phi}_{\psi}(t_i).
\label{creation-operator-quantum-circuit}
\end{equation}
%\begin{align}
%\bar{a}_{\psi}^{\dagger}&=
%\sum_{i=1}^N\sum_{j=1}^Sa^d\delta_t
%\psi(t_i,\mathbf{x}_j)
%e^{iHt_i}\hat{\phi}(\mathbf{x}_j)e^{-iHt_i}\nonumber\\
%&=\sum_{i=1}^N\hat{\phi}_{\psi}(t_i).
%\label{creation-operator-quantum-circuit}
%\end{align} %twocolumn
We work in the field basis \cite{Jordan-Lee-Preskill_1111,Jordan-Lee-Preskill_1112,Klco-Savage_1808},
where the operator
$\hat{\phi}(\mathbf{x})$ is diagonal.
If $k$ qubits
are dedicated to the lattice
site $\mathbf{x}$, then $\hat{\phi}(\mathbf{x})$
is implemented by a linear combination of
$Z$ Pauli matrices,
\begin{equation}
\hat{\phi}(\mathbf{x})=\frac{\phi_{\text{max}}}{2^k-1}
\sum_{i=0}^{k-1}2^i\sigma_{(\mathbf{x},i)}^z.
\label{discretized-phi}
\end{equation}
The effects of truncating and discretizing the spectrum of
$\hat{\phi}(\mathbf{x})$ were studied
in Refs.~\cite{Jordan-Lee-Preskill_1111,Jordan-Lee-Preskill_1112,Klco-Savage_1808}, where it is also shown that it is enough
to take $k$ quite small in a broad range of situations.
The operator~\eqref{discretized-phi} is not unitary
but can be implemented
with linear combination of unitaries (LCU)
\cite{Childs-Kothari-Somma_1511}.
Furthermore, the operator $\bar{a}_{\psi}^{\dagger}$ can
be implemented with LCU as well, with probability
of success
\begin{equation} %onecolumn
\rho=
\bigg(\frac{\norm{\bar{a}_{\psi}^{\dagger}\ket\Omega}}{\alpha}\bigg)^2,
\qquad
\alpha=\phi_{\text{max}}
\sum_{i=1}^N\sum_{j=1}^Sa^d\delta_t
\abs{\psi(t_i,\mathbf{x}_j)}.
\label{probability-denominator}
\end{equation}
%\begin{align}
%\rho&=
%\bigg(\frac{||\bar{a}_{\psi}^{\dagger}\ket\Omega||}{\alpha}\bigg)^2,\\
%\alpha&=\phi_{\text{max}}
%\sum_{i=1}^N\sum_{j=1}^Sa^d\delta_t
%\abs{\psi(t_i,\mathbf{x}_j)}.
%\label{probability-denominator}
%\end{align} %twocolumn
It is not easy to determine
$\rho$ exactly (more is said on this at the end of this section),
but one could use, for example,
the techniques described
in Ref.~\cite{Brassard-Hoyer-Mosca-Tapp_00} to find
numerical estimates for it. Then one would have
to repeat state preparation
$O(1/\rho)$ times to get the desired
initial state, or, alternatively, one could apply
amplitude amplification
\cite{Brassard-Hoyer-Mosca-Tapp_00} to obtain
a quadratic speedup at the
expense of a larger circuit depth.
Typically, we need to prepare two incoming wavepackets.
If we denote by $\rho_1$ and $\rho_2$ the probabilities
of obtaining the two wavepackets, the total probability is
given by $\rho=\rho_1\rho_2$. Generalization
to more than two wavepackets is straightforward.

Before moving on, we remark on how
to implement the
bound-state creation operator~\eqref{bound-state-creation}.
The discussion goes along the line described so far,
except that now, instead of~\eqref{discretized-phi},
we have to use the operator
\begin{equation}
\hat{\phi}(\mathbf{x})^2=
\Big(\frac{\phi_{\text{max}}}{2^k-1}\Big)^2
\sum_{i,j=0}^{k-1}2^{i+j}\sigma_{(\mathbf{x},i)}^z
\sigma_{(\mathbf{x},j)}^z
\label{discretized-squared-phi}
\end{equation}
up to a shift. We are free to choose this shift
in such a way as to eliminate the terms
with $i=j$ in~\eqref{discretized-squared-phi},
which are proportional
to $\openone$. In this way, instead of
$\alpha$ in~\eqref{probability-denominator},
we obtain
\begin{equation}
\alpha_{\text{b}}\approx\frac{2}{3}\phi_{\text{max}}^2
\sum_{i=1}^N\sum_{j=1}^Sa^d\delta_t
\abs{\psi(t_i,\mathbf{x}_j)}.
\end{equation}

\subparagraph{Circuit description}
We want to give a high-level description of a
circuit implementing $\bar{a}_{\psi}^{\dagger}$, so we focus
only on the dependence on the lattice sites and
the time to keep the discussion concise.
We take a register of $N_a=\ceil{\log(kNS)}$
ancillary qubits and we label the
computational basis as
$\ket{t_i,\mathbf{x}_j}$, with $i=1,\dots,N$
and $j=1,\dots,S$.

We define operators $V_{\psi}$ and $V'_{\psi}$ such
that
\begin{align}
V_{\psi}\ket{0}^{\otimes N_a}&=
\frac{1}{\sqrt{\norm{\psi}_1}}
\sum_{i=1}^N\sum_{j=1}^S
\sqrt{a^d\delta_t\psi(t_i,\mathbf{x}_j)}
\ket{t_i,\mathbf{x}_j},\\
{V'}^{\dagger}_{\psi}\ket{0}^{\otimes N_a}&=
\frac{1}{\sqrt{\norm{\psi}_1}}
\sum_{i=1}^N\sum_{j=1}^S
\Big(\sqrt{a^d\delta_t\psi(t_i,\mathbf{x}_j)}\Big)^*
\ket{t_i,\mathbf{x}_j}.
\end{align}
Then the circuits in Fig.~\ref{quantum-circuit}
implement the operator~$\bar{a}_{\psi}^{\dagger}$ written
in~\eqref{creation-operator-quantum-circuit}
up to a normalization factor and when
the state $\ket{0}^{\otimes N_a}$ is obtained
by measurement of the ancillary register.
In general we have
$t_1<0$, $t_N>0$ and $t_i-t_{i+1}=-\delta_t$; therefore,
the sequence of time evolution operators appearing
in~\hyperref[circuit1]{\textbf{Circuit 1}}
consists of a backward evolution
for time $\abs{t_1}$, followed by $N$ steps
forward, each one of time $\delta_t$
for a total of $t_N-t_1$, and by
a final backward evolution for time $t_N$.
In~\hyperref[circuit2]{\textbf{Circuit 2}}
we notice that controlling only each
$\hat{\phi}$ is equivalent to controlling
$e^{-iHt_i}\hat{\phi}e^{iHt_i}$
because $e^{-iHt_i}e^{iHt_i}=\openone$.
To prepare two wavepackets, we can take
two ancillary registers, apply in sequence
the circuit of Fig.~\ref{quantum-circuit},
one for each wavepacket, and measure both ancillary registers
at the end.

\subparagraph{Complexity}
We show that the complexity
of~\hyperref[circuit1]{\textbf{Circuit 1}}
is dominated by the sequence of time evolutions.
We do not want to discuss here how to implement
the time evolution, as this is not in the scope
of this work. Moreover, we take $k=\tilde{O}(1)$.

The sequence of operators $\Phi(t_1),\dots,\Phi(t_N)$
requires $O(kSN)=\tilde{O}(SN)$ gates, as well as the
operators $V$ and $V'$, which basically provide
generic state preparation on $N_a$ qubits,
and we have $S<\mathcal{V}$.
We can estimate the error introduced by discretizing
the integral over $t$, and hence how large $N$
needs to be, in the following way. We
take $t_0=t_1-\delta_t/2$ and $T=t_N-t_1+\delta_t$.
Then we split the integral from $t_0$ to
$t_0+T$ into $N$ integrals in the following way:
\begin{align} %onecolumn
\int_{t_0}^{t_0+T}\!\psi(t,\mathbf{x})
e^{iHt}\hat{\phi}(\mathbf{x})e^{-iHt}dt&=
\sum_{i=1}^N
\int_{t_i-\frac{\delta_t}{2}}^{t_i+\frac{\delta_t}{2}}\!
\psi(t,\mathbf{x})
e^{iHt}\hat{\phi}(\mathbf{x})e^{-iHt}dt\nonumber\\
&=\sum_{i=1}^N
\int_{-\frac{\delta_t}{2}}^{\frac{\delta_t}{2}}\!
\psi(t_i+t,\mathbf{x})
e^{iHt_i}
e^{iHt}\hat{\phi}(\mathbf{x})e^{-iHt}
e^{-iHt_i}dt,
\end{align}
%\begin{align}
%&\int_{t_0}^{t_0+T}\!\psi(t,\mathbf{x})
%e^{iHt}\hat{\phi}(\mathbf{x})e^{-iHt}dt=\nonumber\\
%&\sum_{i=1}^N
%\int_{t_i-\frac{\delta_t}{2}}^{t_i+\frac{\delta_t}{2}}\!
%\psi(t,\mathbf{x})
%e^{iHt}\hat{\phi}(\mathbf{x})e^{-iHt}dt=\nonumber\\
%&\sum_{i=1}^N
%\int_{-\frac{\delta_t}{2}}^{\frac{\delta_t}{2}}\!
%\psi(t_i+t,\mathbf{x})
%e^{iHt_i}
%e^{iHt}\hat{\phi}(\mathbf{x})e^{-iHt}
%e^{-iHt_i}dt,
%\end{align} %twocolumn
where from the
first line to the second %onecolumn
%second line to the third %twocolumn
line we have shifted
the integration variable, $t\to t+t_i$.
We expand $e^{iHt}\hat{\phi}(\mathbf{x})e^{-iHt}$
using the formula
\begin{equation}
e^ABe^{-A}=B+[A,B]+\frac{1}{2}\big[A,[A,B]\big]+\dots
\end{equation}
and we expand
$\psi(t_i+t,\mathbf{x})$ using
the Taylor expansion around $t=0$ up to order
$t^2$. Odd orders in $t$ do not contribute
because the integration domain is symmetric around
zero. The leading order gives us exactly
the operators appearing
in~\eqref{creation-operator-quantum-circuit}.
We use the spectral norm of the next
to leading order to estimate the error due
to discretization and we apply the triangular inequality:
%\begin{widetext} %switch
\begin{align}
\epsilon=&
\frac{\delta_t^3}{24}
\norm{\sum_{i=1}^N
e^{iHt_i}
\Big(
\ddot{\psi}(t_i,\mathbf{x})\hat{\phi}(\mathbf{x})+
2i\dot{\psi}(t_i,\mathbf{x})
[H,\hat{\phi}(\mathbf{x})]-
\psi(t_i,\mathbf{x})
\big[H,[H,\hat{\phi}(\mathbf{x})]\big]
\Big)
e^{-iHt_i}}\nonumber\\
\le&\frac{\delta_t^2}{24}
\Big(
\sum_{i=1}^N\delta_t
|\ddot{\psi}(t_i,\mathbf{x})|
\norm{\hat{\phi}(\mathbf{x})}+
2\sum_{i=1}^N\delta_t|\dot{\psi}(t_i,\mathbf{x})|
\norm{[H,\hat{\phi}(\mathbf{x})]}+\sum_{i=1}^N\delta_t|\psi(t_i,\mathbf{x})|
\norm{\big[H,[H,\hat{\phi}(\mathbf{x})]\big]}
\Big).
\end{align}
%\end{widetext} %switch
The dominant contribution is given by the term
with $\big[H,[H,\hat{\phi}(\mathbf{x})]\big]$, and
the quantity
\begin{equation}
\sum_{i=1}^N\delta_t|\psi(t_i,\mathbf{x})|
\end{equation}
is approximately a constant independent of the lattice
and the precision. Finally, given that
$T=\delta_tN$, we have
\begin{equation}
N=\tilde{O}\bigg(\frac{T}{\sqrt{\epsilon}}
\sqrt{\norm{\big[H,[H,\hat{\phi}(\mathbf{x})]\big]}}
\bigg).
\end{equation}
If we use a first-order Suzuki-Trotter
formula to implement the time evolution, we
need~\cite{Childs-Su-Tran-Wiebe-Zhu_1912}
\begin{equation}
N_{\text{ST}}=O\bigg(\frac{T^2}{\epsilon}
\norm{[H_{\phi},H_{\pi}]}\bigg)
\end{equation}
Trotter steps to keep the error below $\epsilon$.
As one may check by looking at
Eqs.~\eqref{commutator-1} and
\eqref{commutator-2},
the commutator $[H_{\phi},H_{\pi}]$
scales with the total size of the lattice,
as it involves a summation over all the sites, while
$\big[H,[H,\hat{\phi}(\mathbf{x})]\big]$
involves only a few neighbors of $\mathbf{x}$. Thus, $N_{\text{ST}}$ clearly dominates
over $N$ even if we consider higher-order product
formulae. This shows that the complexity is
determined by the time evolution.

%\begin{widetext} %switch
For the $\phi^4$ theory we have
\begin{equation}
\big[H,[H,\hat{\phi}(\mathbf{x})]\big]=
\frac{1}{a^2}\sum_{i=1}^d\Big[
2\hat{\phi}(\mathbf{x})-
\hat{\phi}(\mathbf{x}+\hat{\mathbf{r}}_i)-
\hat{\phi}(\mathbf{x}-\hat{\mathbf{r}}_i)
\Big]+m_0^2\hat{\phi}(\mathbf{x})+
\frac{\lambda_0}{3!}\hat{\phi}(\mathbf{x})^3,
\label{commutator-1}
\end{equation}
\begin{multline}
[H_{\phi},H_{\pi}]=i\sum_{\mathbf{x}}a^d\bigg[
\frac{1}{a^2}\sum_{j=1}^d\bigg(
\hat{\phi}(\mathbf{x})\hat{\pi}(\mathbf{x})+
\hat{\phi}(\mathbf{x}+\hat{\mathbf{r}}_j)
\hat{\pi}(\mathbf{x}+\hat{\mathbf{r}}_j)-
\hat{\phi}(\mathbf{x}+\hat{\mathbf{r}}_j)
\hat{\pi}(\mathbf{x})-
\hat{\phi}(\mathbf{x})
\hat{\pi}(\mathbf{x}+\hat{\mathbf{r}}_j)-
\frac{i}{a^d}\bigg)+\\
+\frac{m_0^2}{2}\bigg(
2\hat{\phi}(\mathbf{x})\hat{\pi}(\mathbf{x})-
\frac{i}{a^d}\bigg)+
\frac{\lambda_0}{12}\bigg(
2\hat{\phi}^3(\mathbf{x})\hat{\pi}(\mathbf{x})-
\frac{3i}{a^d}\hat{\phi}(\mathbf{x})^2\bigg)
\bigg].
\label{commutator-2}
\end{multline}
%\end{widetext} %switch

\subparagraph{On the success probability}
Here we want to show what to expect regarding the
probability of success. We write it again here:
\begin{gather} %onecolumn
\rho=
\bigg(\frac{||\bar{a}_{\psi}^{\dagger}\ket{\Omega}||}{\alpha}\bigg)^2,\\
\alpha=\phi_{\text{max}}
\sum_{i=1}^N\sum_{j=1}^Sa^d\delta_t
\abs{\psi(t_i,\mathbf{x}_j)},\qquad\qquad
\bar{a}_{\psi}^{\dagger}\ket{\Omega}=
\sum_{i=1}^N\sum_{j=1}^Sa^d\delta_t
\psi(t_i,\mathbf{x}_j)
e^{it_iH}\hat{\phi}(\mathbf{x}_j)\ket{\Omega}.
\end{gather}
%\begin{align}
%\rho&=
%\bigg(\frac{||\bar{a}_{\psi}^{\dagger}\ket{\Omega}||}{\alpha}\bigg)^2,\\
%\alpha&=\phi_{\text{max}}
%\sum_{i=1}^N\sum_{j=1}^Sa^d\delta_t
%\abs{\psi(t_i,\mathbf{x}_j)},\label{alpha}\\
%\bar{a}_{\psi}^{\dagger}\ket{\Omega}&=
%\sum_{i=1}^N\sum_{j=1}^Sa^d\delta_t
%\psi(t_i,\mathbf{x}_j)
%e^{it_iH}\hat{\phi}(\mathbf{x}_j)\ket{\Omega}.
%\end{align} %twocolumn
Near the continuum limit, $||\bar{a}_{\psi}^{\dagger}\ket{\Omega}||^2$
should approach its continuum value
\begin{equation}
Z\int\!d^dp\,\frac{|\tilde{g}(\mathbf{p})
\tilde{f}_1(E(\mathbf{p}),\mathbf{p})|^2}{2E(\mathbf{p})},
\end{equation}
where $Z$ depends on the normalization of the field
operator $\hat{\phi}(\mathbf{x})$.
The summation
in $\alpha$ can be approximated by an integral, so we have
\begin{equation}
\rho\sim\frac{Z}{\phi_{\text{max}}^2}
\frac{\int\!d^dp\,\frac{|\tilde{g}(\mathbf{p})
\tilde{f}_1(E(\mathbf{p}),\mathbf{p})|^2}{2E(\mathbf{p})}}{\big(
\int\!d^Dx\,|\psi(x)|\big)^2}.
\end{equation}

Since $\tilde{g}(\mathbf{p})\tilde{f}_1(p_0,\mathbf{p})$
has support of size $\delta_E$ in the $p_0$
direction, and support of size $\delta_p$ in the other directions,
we have
\begin{equation} %onecolumn
\bigg(\int\!d^Dx\,|\psi(x)|\bigg)^2\sim
\delta_E^{-1}\delta_p^d\int\!d^Dx\,|\psi(x)|^2=
\delta_E^{-1}\delta_p^d\int\!d^Dp\,
\Big|\tilde{g}(\mathbf{p})\tilde{f}_1(p)
\frac{p_0+E(\mathbf{p})}{2E(\mathbf{p})}\Big|^2.
\end{equation}
%\begin{align}
%\bigg(\int\!\!d^Dx\,|\psi(x)|\bigg)^2&\sim
%\frac{1}{\delta_E\delta_p^d}
%\int\!\!d^Dx\,|\psi(x)|^2\notag\\&
%=\frac{1}{\delta_E\delta_p^d}\int\!\!d^Dp\,
%\Big|\tilde{g}(\mathbf{p})\tilde{f}_1(p)
%\frac{p_0+E(\mathbf{p})}{2E(\mathbf{p})}\Big|^2.
%\end{align} %twocolumn
In the limit $\delta_p\to0$ we can approximate
$\tilde{g}$ with a delta function centered in $\bar{\mathbf{p}}$,
and recalling $\bar{E}=E(\bar{\mathbf{p}})$, we obtain
\begin{equation} %onecolumn
\frac{\int\!d^dp\,\frac{|\tilde{g}(\mathbf{p})
\tilde{f}_1(E(\mathbf{p}),\mathbf{p})|^2}{
2E(\mathbf{p})}}{\big(\int\!d^Dx\,|\psi(x)|\big)^2}\sim
\frac{\delta_p^d}{\bar{E}}
\frac{\delta_E|\tilde{f}_1(\bar{E},\bar{\mathbf{p}})|^2}{
\int_{-\infty}^{+\infty}\!dp_0\,|\tilde{f}_1(p_0,\bar{\mathbf{p}})
\frac{p_0+\bar{E}}{2\bar{E}}|^2}
\sim\frac{\delta_p^d}{\bar{E}}
\end{equation}
%\begin{align}
%\frac{\int\!d^dp\,\frac{|\tilde{g}(\mathbf{p})
%\tilde{f}_1(E(\mathbf{p}),\mathbf{p})|^2}{
%2E(\mathbf{p})}}{\big(\int\!d^Dx\,|\psi(x)|\big)^2}&\sim
%\frac{\delta_p^d}{\bar{E}}
%\frac{\delta_E|\tilde{f}_1(\bar{E},\bar{\mathbf{p}})|^2}{
%\int_{-\infty}^{+\infty}\!dp_0\,|\tilde{f}_1(p_0,\bar{\mathbf{p}})
%\frac{p_0+\bar{E}}{2\bar{E}}|^2}\notag\\
%\sim\frac{\delta_p^d}{\bar{E}}
%\end{align} %twocolumn
and
\begin{equation}
\rho=O
\bigg(\frac{Z}{\phi_{\text{max}}^2}
\frac{\delta_p^d}{\bar{E}}\bigg).
\end{equation}

We now turn our attention to the factor $Z/\phi_{\text{max}}^2$.
Set
\begin{equation} %onecolumn
\sigma_E=\sqrt{\bra{\psi_E}\hat{\phi}(0)^2\ket{\psi_E}},
\qquad
\sigma_0=\sqrt{\bra{\Omega}\hat{\phi}(0)^2\ket{\Omega}},
\end{equation}
%\begin{align}
%\sigma_E&=\sqrt{\bra{\psi_E}\hat{\phi}(0)^2\ket{\psi_E}},\\
%\sigma_0&=\sqrt{\bra{\Omega}\hat{\phi}(0)^2\ket{\Omega}},
%\end{align} %twocolumn
where $\ket{\psi_E}$ is a state with maximum energy $E$ and
$E$ is the energy of the process being simulated.
In Ref.~\cite{Jordan-Lee-Preskill_1112}, as a
consequence of Chebyshev's inequality, it is shown that
\begin{equation}
\phi_{\text{max}}=O\bigg(
\sigma_E\sqrt{\frac{\mathcal{V}}{\epsilon_{\text{trunc}}}}
\bigg),
\label{truncation-bound-Preskill}
\end{equation}
where $\epsilon_{\text{trunc}}$
is the error due to truncation
of the spectrum of $\hat{\phi}(\mathbf{x})$.
In the field representation we have
\begin{equation}
\ket{\psi_E}=\int_{-\infty}^{+\infty}\!\!d\phi_1\dots
\int_{-\infty}^{+\infty}\!\!d\phi_{\mathcal{V}}\,
\psi_E(\phi_1,\dots,\phi_{\mathcal{V}})
\ket{\phi_1\dots\phi_{\mathcal{V}}},
\label{field-representation}
\end{equation}
where $1,\dots,\mathcal{V}$ is some
labeling of the lattice sites.
It is well known~\cite{Manousakis} that
$\psi_E(\phi_1,\dots,\phi_{\mathcal{V}})$
decays rapidly (exponentially or more)
outside the classically forbidden region
$V(\phi_1,\dots,\phi_{\mathcal{V}})>E$, where
$V(\phi_1,\dots,\phi_{\mathcal{V}})$ is the potential
associated with the Hamiltonian $H$.
From the discussion in the Appendix it is clear
that we can improve the
bound~\eqref{truncation-bound-Preskill} to
\begin{equation}
\phi_{\text{max}}=O\bigg(
\sigma_E\sqrt{\mathcal{V}}
+\log\frac{\mathcal{V}}{\epsilon_{\text{trunc}}}\bigg).
\label{new-bound}
\end{equation}

Moreover, from the analysis
in Ref.~\cite{Jordan-Lee-Preskill_1112} it follows that
(at most) $\sigma_E=O(\sqrt{E}\sigma_0)$,
and thus, ignoring the logarithmic contribution, we have
\begin{equation}
\phi_{\text{max}}=
O\big(\sigma_0\sqrt{E\mathcal{V}}\big).
\end{equation}
We can use the Kållen-Lehmann representation
of the two-point function in the continuum to estimate
$\sigma_0$.
We have
\begin{equation} %onecolumn
\bra{\Omega}\hat{\phi}(x)\hat{\phi}(y)\ket{\Omega}=
Z\Delta_+(x-y,m^2)+
\int_{4m^2}^{+\infty}\!\rho(\mu)\Delta_+(x-y,\mu)d\mu,
\label{spectral-representation}
\end{equation}
%\begin{multline}
%\bra{\Omega}\hat{\phi}(x)\hat{\phi}(y)\ket{\Omega}=\\
%Z\Delta_+(x-y,m^2)+
%\int_{4m^2}^{+\infty}\!\rho(\mu)\Delta_+(x-y,\mu)d\mu,
%\label{spectral-representation}
%\end{multline} %twocolumn
where $\rho(\mu)\ge0$ is the so-called spectral function
and
\begin{equation}
\Delta_+(x,\mu)=\frac{1}{(2\pi)^d}\int\!
\frac{d^dp}{2\sqrt{|\mathbf{p}|^2+\mu}}e^{-ip\cdot x}.
\end{equation}
In renormalizable models such as $\phi^4$, the spectral function
is known to decay as $1/\mu$ at all orders
of perturbation theory, ensuring the convergence
of the integral in~\eqref{spectral-representation}.
Therefore, we may assume that,
at coincident points $x=y$, the behavior
of the two-point function is determined by
$\Delta_+(0,m^2)$, which diverges logarithmically
for $d=1$ and linearly and quadratically (apart from
logarithmic factors) for $d=2$ and $d=3$, respectively. As
$1/a$ is a natural cutoff on the lattice, we may conclude
that
\begin{equation}
\phi_{\text{max}}=\tilde{O}\bigg(
\sqrt{\frac{EZ\mathcal{V}}{a^{d-1}}}\bigg),
\end{equation}
where we use the notation $\tilde{O}(\cdot)=
O(\cdot\polylog(\cdot))$,
and consequently
\begin{equation}
\rho\gtrsim\frac{a^{d-1}\delta_p^d}{\mathcal{V}E\bar{E}},
\label{success-probability}
\end{equation}
up to logarithmic factors.
As we are interested in the continuum limit,
the dominant factor is $a^{d-1}/\mathcal{V}$.
From the discussion in the Appendix we expect
this bound to be quite loose. For the perturbative
$\phi^4$ theory, for instance, we can replace
$\mathcal{V}$ with $\sqrt{\mathcal{V}}$.

\section{Comparison with the JLP algorithm}
\label{scattering-protocol}
In this section we compare the complexity
scaling of our strategy with the complexity scaling of the JLP
strategy~\cite{Jordan-Lee-Preskill_1111,Jordan-Lee-Preskill_1112}
in the case when both methods are applicable,
that is when we want to prepare elementary particles.
We consider only the $\phi^4$ theory.
The settings in the two cases are very similar and a comparison is immediate.

The JLP algorithm
can be summarized in five steps:
\begin{enumerate}
\item Free vacuum preparation
\item Creation of free wavepackets
\item Adiabatic transformation of the free wavepackets
into interacting wavepackets
\item Time evolution;
\item Measurements
\end{enumerate}
As stated in the introduction, our strategy requires the
preparation of the interacting vacuum. While other
techniques, such as variational approaches, might be more
suitable in practice, to study the complexity we choose
here to do the following:
\begin{enumerate}
\item Free vacuum preparation
\item Adiabatic transformation of the free vacuum into
the interacting vacuum
\item Creation of interacting wavepackets
\item Time evolution
\item Measurements
\end{enumerate}
Steps 4 and 5 are the same in the two approaches,
and in both cases the bottleneck of complexity is
in the initial-state preparation, steps 1,2 and 3,
so we focus on these steps only.
We consider here the case of two particles in the initial state.
As the complexity depends on the success probability,
simulating scattering between three or more incoming particles
becomes more and more inefficient in our approach,
and is of little or no practical interest in general.
We assume the reader
is familiar with
Ref.~\cite{Jordan-Lee-Preskill_1112},
especially with Secs. 3.2 and 4.2. We do not
want to discuss the details of how time evolution
is implemented, although it may have an impact
on complexity. Our goal is to give a rough estimate,
certainly not exhaustive,
and to compare it with the results obtained by Jordan, Lee, and Preskill.
To this end and following their approach,
we assume we can implement time evolution
with a $k$th-order Suzuki-Trotter formula with large $k$.
The gate cost for time $t$ on a lattice of $\mathcal{V}$
sites is
\begin{equation}
(\mathcal{V}t)^{1+\frac{1}{2k}},
\end{equation}
which we indicate as
\begin{equation}
(\mathcal{V}t)^{1+o(1)}
\label{Suzuki-Trotter-scaling}
\end{equation}
to simplify the presentation. The little-$o$ notation
is used also to include logarithmic factors
implicit in the $\tilde{O}$ notation.

The most-time-consuming
parts are typically the free vacuum preparation and
the adiabatic transformation.
The complexity for the free vacuum preparation,
using the Kitaev-Webb
method~\cite{Kitaev-Webb_0801,Bauer-Deliyannis-Freytsis-Nachman_2109},
is $O(\mathcal{V}^{2.376})$. This exponent is determined
by the classical computation of the LDL decomposition
of the covariance matrix. The quantum circuit has a depth
of $\tilde{O}(\mathcal{V}^2)$ and requires $O(\log\mathcal{V})$
ancillary qubits~\cite{Bauer-Deliyannis-Freytsis-Nachman_2109}.

The adiabatic transformation of the wavepackets
requires a modification with respect to a traditional
treatment to take into account the
fact that wavepackets are not eigenstates of the Hamiltonian,
as described in Secs.~3.1 and 4.2
of~\cite{Jordan-Lee-Preskill_1112}. In particular,
the adiabatic transformation, of total time
$\tau_{\text{wp}}$,
needs to be split into $J\sim\sqrt{\tau_{\text{wp}}}$
steps and interspersed with
backward time evolution to suppress the dynamical phases.
This causes the adiabatic error $\epsilon$ to vanish like
$J^2/\tau_{\text{wp}}^2\sim1/\tau_{\text{wp}}$ instead of
as $1/\tau_{\text{wp}}^2$. Our approach has the
advantage that the adiabatic transformation
is performed on the vacuum, rather than on the wavepackets,
and there is no need to suppress the dynamical phases.
To continue the discussion, we distinguish between the
weak-coupling regime and the strong-coupling regime from now on.
In both regimes we take the adiabatic paths chosen
in Refs.~\cite{Jordan-Lee-Preskill_1111,Jordan-Lee-Preskill_1112}.

\subparagraph{Weak coupling}
We consider scaling in the continuum limit.
To determine the scaling of
the time $\tau_{\text{vac}}$ required to perform the adiabatic
transformation in the vacuum, we can use the analysis
of adiabaticity
in Sec. 4.2 in Ref.~\cite{Jordan-Lee-Preskill_1112},
taking $J=1$. Setting $V=a^d\mathcal{V}$, this leads to
\begin{equation}
\tau_{\text{vac}}=
\begin{cases*}
\tilde{O}\Big(\sqrt{\frac{V}{\epsilon}}\Big)&$d=1,2$,\\
\tilde{O}\Big(\sqrt{\frac{V}{a^4\epsilon}}\Big)&$d=3$,
\end{cases*}
\end{equation}
to be compared with the result obtained by Jordan, Lee and Preskill
\begin{equation}
\tau_{\text{wp}}=
\begin{cases*}
\tilde{O}\Big(\frac{V}{\epsilon}\Big)&$d=1,2$,\\
\tilde{O}\Big(\frac{V}{a^6\epsilon}\Big)&$d=3$.
\end{cases*}
\end{equation}

We next consider the number of gates needed for the
free vacuum preparation, $G_{\text{prep}}$,
and the two kinds of adiabatic transformation,
$G_{\text{wp}}$ and $G_{\text{vac}}$.
To this end, we take $a\sim\sqrt{\epsilon}$
and $V\sim\log(1/\epsilon)$, as argued
in Ref.~\cite{Jordan-Lee-Preskill_1112}. There,
in section 3.2,
they also provide $G_{\text{prep}}=\tilde{O}(1/\epsilon^{1.188d})$
and, by~\eqref{Suzuki-Trotter-scaling},
\begin{equation}
G_{\text{wp}}\sim
\begin{cases*}
\Big(\frac{1}{\epsilon}\Big)^{1.5+o(1)}&$d=1$,\\
\Big(\frac{1}{\epsilon}\Big)^{2+o(1)}&$d=2$,\\
\Big(\frac{1}{\epsilon}\Big)^{5.5+o(1)}&$d=3$.\\
\end{cases*}
\end{equation}
Following the same reasoning we find
\begin{equation}
G_{\text{vac}}\sim
\begin{cases*}
\Big(\frac{1}{\epsilon}\Big)^{1+o(1)}&$d=1$,\\
\Big(\frac{1}{\epsilon}\Big)^{1.5+o(1)}&$d=2$,\\
\Big(\frac{1}{\epsilon}\Big)^{3+o(1)}&$d=3$.\\
\end{cases*}
\end{equation}
We see that $G_{\text{prep}}$ dominates
over $G_{\text{vac}}$ for all the dimensions.

The advantage obtained by avoiding the adiabatic
transformation on wavepackets
is partially spoiled by the fact
that the wavepacket creation does not succeed with
a probability of 1. For simplicity, we consider two similar
wavepackets, so, by equation~\eqref{success-probability} (replacing
$\mathcal{V}$ with $\sqrt{\mathcal{V}}$ in weakly coupled
$\phi^4$ theory), we find that the probability goes as
\begin{equation}
\rho\gtrsim\bigg(\frac{a^{d-1}\delta_p^d}{\sqrt{\mathcal{V}}E\bar{E}}
\bigg)^2.
\end{equation}
This means that we need to repeat
the state preparation
\begin{equation}
O\bigg(\frac{1}{\rho}\bigg)=\tilde{O}\bigg(\frac{V}{a^{3d-2}}\bigg)
\end{equation}
times to obtain the correct initial state.
Notice that the LDL decomposition of the covariance
matrix does not need to be repeated every time, so the
total complexity of the state preparation protocol
proposed here is obtained by multiplying the
depth of the free vacuum preparation by $1/\rho$,
which gives
\begin{equation}
G=\tilde{O}\Big(\frac{\mathcal{V}^2}{\rho}\Big)\sim
\begin{cases*}
\Big(\frac{1}{\epsilon}\Big)^{1.5+o(1)}&$d=1$,\\
\Big(\frac{1}{\epsilon}\Big)^{4+o(1)}&$d=2$,\\
\Big(\frac{1}{\epsilon}\Big)^{6.5+o(1)}&$d=3$,\\
\end{cases*}
\end{equation}
with depth given by $\tilde{O}(\mathcal{V}^2)=\tilde{O}(1/\epsilon^d)$.
The total complexity can be improved at the expense of
a larger depth by use of amplitude amplification, in which case
we have
\begin{equation}
G'=\tilde{O}\bigg(\frac{\mathcal{V}^2}{\sqrt{\rho}}\bigg)\sim
\begin{cases*}
\Big(\frac{1}{\epsilon}\Big)^{1.25+o(1)}&$d=1$,\\
\Big(\frac{1}{\epsilon}\Big)^{3+o(1)}&$d=2$,\\
\Big(\frac{1}{\epsilon}\Big)^{4.75+o(1)}&$d=3$.\\
\end{cases*}
\end{equation}
For a direct comparison, the total complexity
of the JLP algorithm is given by
\begin{equation}
G_{\text{JLP}}\sim
\begin{cases*}
\Big(\frac{1}{\epsilon}\Big)^{1.5+o(1)}&$d=1$,\\
\Big(\frac{1}{\epsilon}\Big)^{2.376+o(1)}&$d=2$,\\
\Big(\frac{1}{\epsilon}\Big)^{5.5+o(1)}&$d=3$.\\
\end{cases*}
\end{equation}

\subparagraph{Strong coupling}
Because of the triviality issue of the $\phi^4$ theory
in a three-dimensional space, we can have strong coupling
only for $d=1,2$, when we approach the phase transition.
For the success probability, we take
\begin{equation}
\rho\gtrsim\bigg(\frac{a^{d-1}\delta_p^d}{\mathcal{V}E\bar{E}}
\bigg)^2.
\label{strong-success-probability}
\end{equation}
We limit our discussion to the
scaling of complexity with the coupling strength and
the momenta of the incoming particle.
If the phase transition occurs at the critical value
$\lambda_{\text{c}}$,
we can take $\abs{\lambda_0-\lambda_{\text{c}}}$
as a measure of the coupling strength. To estimate
the scaling of the adiabatic time
with this quantity, we use the results
in Ref.~\cite{Jansen-Ruskai-Seiler_0603}, implying
$\tau_{\text{vac}}\sim1/m^3$.
The temporal
size $T$ of the wavepacket $\psi$
in~\eqref{creation-operator-quantum-circuit},
which determines the duration of the time
evolution required to create the wavepacket in our approach,
grows at most as $1/m$, by the uncertainty principle,
so $\tau_{\text{vac}}$ dominates over $T$.
Near the phase transition,
the physical mass vanishes as
\begin{equation}
m\sim
\begin{cases*}
\lambda_{\text{c}}-\lambda_0&$d=1$,\\
(\lambda_{\text{c}}-\lambda_0)^{0.63}&$d=2$,
\end{cases*}
\end{equation}
which gives
\begin{equation}
\tau_{\text{vac}}\sim
\begin{cases*}
\Big(\frac{1}{\lambda_{\text{c}}-\lambda_0}\Big)^3&$d=1$,\\
\Big(\frac{1}{\lambda_{\text{c}}-\lambda_0}\Big)^{1.89}&$d=2$.
\end{cases*}
\end{equation}
Furthermore,
the probability of success~\eqref{strong-success-probability}
does not depend explicitly on the mass gap. However,
the volume has to be large enough to contain
the wavepackets, which in turn have linear size
proportional to $1/m$; hence,
\begin{equation}
V\sim
\begin{cases*}
\frac{1}{\lambda_{\text{c}}-\lambda_0}&$d=1$,\\
\Big(\frac{1}{\lambda_{\text{c}}-\lambda_0}\Big)^{1.26}&$d=2$.
\end{cases*}
\end{equation}
Considering the volume dependence
in equation~\eqref{Suzuki-Trotter-scaling},
the adiabatic transformation has a stronger
scaling than the free vacuum preparation.
Taking into account the success
probability~\eqref{strong-success-probability},
we find that, at fixed $a$ and
incoming momenta, the total complexity scales
with the coupling strength as
\begin{equation}
G_{\text{strong}}\sim
\begin{cases*}
\bigg(\frac{1}{\lambda_{\text{c}}-\lambda_0}\bigg)^{6+o(1)}&$d=1$,\\
\bigg(\frac{1}{(\lambda_{\text{c}}-\lambda_0)}\bigg)^{5.67+o(1)}&$d=2$.
\end{cases*}
\end{equation}
Using amplitude amplification, we have
\begin{equation}
G_{\text{strong}}'\sim
\begin{cases*}
\bigg(\frac{1}{\lambda_{\text{c}}-\lambda_0}\bigg)^{5+o(1)}&$d=1$,\\
\bigg(\frac{1}{(\lambda_{\text{c}}-\lambda_0)}\bigg)^{4.41+o(1)}&$d=2$.
\end{cases*}
\end{equation}
The corresponding result found
in Refs.~\cite{Jordan-Lee-Preskill_1111,Jordan-Lee-Preskill_1112}
is
\begin{equation}
G_{\text{strong,JLP}}\sim
\begin{cases*}
\bigg(\frac{1}{\lambda_{\text{c}}-\lambda_0}\bigg)^{9+o(1)}&$d=1$,\\
\bigg(\frac{1}{(\lambda_{\text{c}}-\lambda_0)}\bigg)^{6.3+o(1)}&$d=2$.
\end{cases*}
\end{equation}

Finally, we consider scaling with incoming momenta, at fixed
coupling strength and volume.
The free vacuum preparation and the adiabatic transformation
in the vacuum do not depend explicitly on the incoming
momenta, but the lattice spacing has
to be small enough to contain momentum mode $p$.
With $a\sim1/p$, the free vacuum preparation has
a cost proportional to $p^{2d}$, while the adiabatic
transformation has a slower growth.
The time evolution required for the wavepacket creation
has a cost count of
$(\mathcal{V}T)^{1+o(1)}\sim p^{d+1+o(1)}$,
since $T$ grows at most linearly with $p$.
The success probability~\eqref{strong-success-probability}
has scaling $p^{4d+2}$. Putting all this together,
we find the scaling
\begin{equation}
G_{\text{strong}}\sim
\begin{cases*}
p^{8+o(1)}&$d=1$,\\
p^{14+o(1)}&$d=2$,
\end{cases*}
\end{equation}
or, using amplitude amplification,
\begin{equation}
G_{\text{strong}}'\sim
\begin{cases*}
p^{5+o(1)}&$d=1$,\\
p^{9+o(1)}&$d=2$.
\end{cases*}
\end{equation}
The corresponding result
in Refs.~\cite{Jordan-Lee-Preskill_1111,Jordan-Lee-Preskill_1112}
is
\begin{equation}
G_{\text{strong,JLP}}\sim
\begin{cases*}
p^{4+o(1)}&$d=1$,\\
p^{6+o(1)}&$d=2$.
\end{cases*}
\end{equation}
The scaling with the spread of the wavepackets in
momentum space, $\delta_p$, can be obtained in
a similar way. In this case, we take $V\sim1/\delta_p^d$,
while we keep $a$ fixed. The total scaling is given by
\begin{equation}
G_{\text{strong}}\sim\bigg(\frac{1}{\delta_p}\bigg)^{6d+o(1)},
\end{equation}
which can be improved to
\begin{equation}
G_{\text{strong}}'\sim\bigg(\frac{1}{\delta_p}\bigg)^{4d+o(1)}
\end{equation}
with the use of amplitude amplification.

\section{Conclusions}
In this paper we provide a quantum algorithm to
create single-particle wavepackets
of a lattice quantum field theory
starting from the vacuum state. The method we propose
is quite general and the idea
is independent of details of
the model. For example, it works equally well for
free and interacting theories. The key aspect
of our strategy is that it is suitable for preparation
of composite particles, which is an important novelty
in the context of quantum simulation of relativistic
scattering. To our knowledge this is the first
work on state preparation of bound states
for digital quantum simulation of scattering.

The work is based on the Haag-Ruelle scattering
theory in the framework of axiomatic quantum
field theory, which is ideal for quantum simulation
as it is developed in the operator formalism.
In this respect, this work also shows the potential
importance that the axiomatic approach might have
for quantum computation applied to quantum field theory,
as both fields are suited
to nonperturbative investigations from first principles.
The Haag asymptotic theorem~\ref{Haag asymptotic theorem}
provides a strong limit to obtain
scattering states, rather than a weak limit
as is the case in the more famous LSZ approach. This feature
is what makes the Haag-Ruelle framework
particularly suitable for quantum simulation,
together with the fact that the convergence rate
of the strong limit is very fast, as discussed
after Theorem~\ref{Haag asymptotic theorem}.

Our present result shows the potential of the
Haag-Ruelle formalism in the context of quantum
simulation. Here we decided to use
LCU because it seems to us
the easiest and most natural route,
but there may be other, more efficient techniques
in the context of digital quantum computation or
in other contexts.
Excluding the steps of vacuum preparation,
and of time evolution and measurement,
our algorithm requires a number of qubits
that grows logarithmically with the size of the wavepacket,
and a circuit depth equivalent to that of the time
evolution. It succeeds with a probability
that vanishes polynomially in the continuum
limit for highly energetic processes
and for narrow wavepackets
in momentum space. The size of the mass gap
is relevant only because the extension
of the wavepacket in the time coordinate
is inversely proportional to the mass gap.

This work decomposes
the problem of state preparation for scattering
into more approachable ones.
On one hand, efficient techniques to prepare
the vacuum state are required. On the other hand,
one has to find interpolating operators with
the right properties for a given particle in
a given theory, and one needs to know the size of
the corresponding lower and upper mass gaps.
On the first front, much work has already been done
in the context of quantum computation,
and we propose a way to prepare the interacting
vacuum by an adiabatic transformation.
We found that a scattering protocol based on
creating wavepackets from the adiabatically prepared
vacuum is in some cases more efficient than the
protocol proposed by Jordan, Lee and
Preskill~\cite{Jordan-Lee-Preskill_1111,Jordan-Lee-Preskill_1112}.
For the second front, standard techniques
of Euclidean lattice field theory are available.
As a next step, the approach of this work
needs to be specialized case by case. Also,
we need to investigate
how gauge invariance and the presence of massless
particles affect this approach.

In the last stages of the present work, we became aware
of two new preprints~\cite{Kreshchuk-Vary-Love_2310,Farrell-Illa-Ciavarella-Savage_2401},
where similar problems
are addressed with different approaches.

\begin{acknowledgments}
We thank Duarte Magano and Miguel
Murça for their introduction to the techniques
of linear combination of unitaries used
in this work. We also thank
Pieralberto Marchetti and Marco Matone for
discussions and help concerning the Haag-Ruelle
scattering theory. Furthermore, we
thank Funda\c{c}\~{a}o para a Ci\^{e}ncia e a Tecnologia
(FCT) (Portugal) for support, namely through project UIDB/04540/2020,
as well as the projects QuantHEP and HQCC
supported by the EU QuantERA ERA-NET Cofund
in Quantum Technologies
and by FCT (QuantERA/0001/2019 and QuantERA/004/2021,
respectively),
and by the EU Quantum Flagship project EuRyQa (Grant No. 101070144).
Finally, Matteo Turco thanks FCT for
support through the grant PRT/BD/154668/2022.
\end{acknowledgments}

\appendix* \section{}
Considering that the probability distribution
determined by $\psi_E$ in~\eqref{field-representation}
is concentrated in the classically allowed region,
the bound~\eqref{truncation-bound-Preskill}
seems loose, and we now explore this idea in more detail.
We take a potential of the form
\begin{gather}
V(\phi_1,\dots,\phi_{\mathcal{V}})=
\tilde{V}(\phi_1,\dots,\phi_{\mathcal{V}})+
\frac{1}{a^2}\sum_{<j,j'>}\big(
\phi_j-\phi_{j'}
\big)^2,\label{potential}\\
\tilde{V}(\phi_1,\dots,\phi_{\mathcal{V}})=
\sum_{j=1}^{\mathcal{V}}v(\phi_j),
\end{gather}
where the summation in~\eqref{potential}
runs over nearest-neighbor pairs on the lattice and
$v(\phi)$ is a polynomial bounded from below
like $\lambda_0\phi^4+m_0^2\phi^2$.
Let $\cube(\phi)$ denote the
$\mathcal{V}$-dimensional hypercube
centered at the origin and of linear size $2\phi>0$.
Let $\phi_{\text{cl}}$ be the smallest $\phi$
such that $\cube(\phi_{\text{cl}})$ contains
the region $\tilde{V}(\phi_1,\dots,\phi_{\mathcal{V}})<E$
($\phi_{\text{cl}}$ would be enough to describe
the entire classical dynamics).
Our aim here is to show that
\begin{equation}
\phi_{\text{max}}=O\bigg(\phi_{\text{cl}}+
\log\Big(\frac{\mathcal{V}}{\epsilon_{\text{trunc}}}\Big)\bigg).
\label{new-truncation-bound}
\end{equation}

The terms in the summation in~\eqref{potential} are
positive semidefinite, which implies that
the classically allowed region of $V$ is contained in the
classical region of $\tilde{V}$ for all energies.
Moreover, as $V\ge\tilde{V}$,
the decay of the wave function outside the classical
region should also be no slower for $V$ than for
$\tilde{V}$~\cite{Gervais-Sakita_77}.
Therefore, for our purposes,
it is enough to consider $\tilde{V}$.

As $\tilde{V}$
does not contain mixing terms between different sites,
we roughly approximate
\begin{align}
\psi_E(\phi_1,\dots,\phi_{\mathcal{V}})\sim
\frac{1}{\sqrt{A}}\psi_1(\phi_1)\cdots\psi_1(\phi_{\mathcal{V}}),
\end{align}
where $\psi_1(\phi_{\text{cl}})=1$ and
$\psi_1(\phi)$ rapidly decreases for
$\phi>\phi_{\text{cl}}$ (for
a quadratic $v$, $\psi_1$ decays like a
Gaussian times a polynomial). We also set
\begin{equation}
u(\phi)=\int_{-\phi}^{\phi}\!|\psi_1(\tilde{\phi})|^2d\tilde{\phi}.
\end{equation}
The normalization constant $A$ is fixed by our requiring
\begin{equation}
\int_{\mathbb{R}^{\mathcal{V}}}\!d^{\mathcal{V}}\phi\,
\frac{1}{A}|\psi_1(\phi_1)\cdots\psi_1(\phi_{\mathcal{V}})|^2=1,
\end{equation}
which gives $A=u(+\infty)^{\mathcal{V}}$. Now we
take $\phi_{\text{max}}>\phi_{\text{cl}}$, and we consider
\begin{equation}
\overline{\cube(\phi_{\text{max}})}=
\mathbb{R}^{\mathcal{V}}\setminus\cube(\phi_{\text{max}}).
\end{equation}
The error due to truncation $\epsilon_{\text{trunc}}$
corresponds to the integral of $|\psi_E|^2$ over
$\overline{\cube(\phi_{\text{max}})}$. Using
\begin{equation}
\int_{\overline{\cube(\phi_{\text{max}})}}=
\int_{\mathbb{R}^{\mathcal{V}}}-\int_{\cube(\phi_{\text{max}})},
\end{equation}
we find
\begin{align} %onecolumn
\epsilon_{\text{trunc}}=
\int_{\overline{\cube(\phi_{\text{max}})}}d^{\mathcal{V}}\phi\,
\frac{1}{A}|\psi_1(\phi_1)\cdots\psi_1(\phi_{\mathcal{V}})|^2
&=1-\bigg[\frac{u(\phi_{\text{max}})}{u(+\infty)}
\bigg]^{\mathcal{V}}\notag\\
&=1-\bigg[1-\frac{u(+\infty)-u(\phi_{\text{max}})}{u(+\infty)}
\bigg]^{\mathcal{V}}
\label{truncation-error}
\end{align}
%\begin{align}
%\epsilon_{\text{trunc}}&=
%\int_{\overline{\cube(\phi_{\text{max}})}}d^{\mathcal{V}}\phi\,
%\frac{1}{A}|\psi_1(\phi_1)\cdots\psi_1(\phi_{\mathcal{V}})|^2
%\notag\\
%&=1-\bigg[\frac{u(\phi_{\text{max}})}{u(+\infty)}
%\bigg]^{\mathcal{V}}\notag\\
%&=1-\bigg[1-\frac{u(+\infty)-u(\phi_{\text{max}})}{u(+\infty)}
%\bigg]^{\mathcal{V}}
%\label{truncation-error}
%\end{align} %twocolumn
From the definition of an exponential,
\begin{equation}
e^{x}\approx\Big(1+\frac{x}{\mathcal{V}}\Big)^{\mathcal{V}},
\end{equation}
it is enough to take
\begin{equation}
\frac{u(+\infty)-u(\phi_{\text{max}})}{u(+\infty)}=
O\Big(\frac{\epsilon_{\text{trunc}}}{\mathcal{V}}\Big)
\end{equation}
to ensure~\eqref{truncation-error} holds.
Since $\psi_1(\phi)$ vanishes exponentially or faster
as $\phi\to+\infty$, so does $u(+\infty)-u(\phi_{\text{max}})$
as $\phi_{\text{max}}\to+\infty$,
and we obtain~\eqref{new-truncation-bound}.

As a next step we would need to determine the relation between
$\phi_{\text{cl}}$ and
$\sigma_E$.
While it is clear that
\begin{equation}
\sigma_E=
O(\phi_{\text{cl}}+\log\mathcal{V}),
\end{equation}
$\phi_{\text{cl}}$ may be asymptotically much larger than
$\sigma_E$
and the form of the potential plays an important role.
To see what $\phi_{\text{cl}}$ looks like,
and to compare~\eqref{new-truncation-bound}
with~\eqref{truncation-bound-Preskill}, we focus
on a more specific example, namely the $\phi^4$ theory
at energy $E=O(1)$. We have
\begin{equation}
v(\phi)=a^d\big(\lambda_0\phi^4+m_0^2\phi^2\big).
\end{equation}
It can be shown that
\begin{align}
\phi_{\text{cl}}&=\sqrt{\frac{
-m_0^2+\sqrt{4a^{-d}\lambda_0E+\mathcal{V}m_0^4}}{2\lambda_0}},&
\qquad&m_0^2<0,\\
\phi_{\text{cl}}&=\sqrt{\frac{
-m_0^2+\sqrt{4a^{-d}\lambda_0E+m_0^4}}{2\lambda_0}},&
\qquad&m_0^2>0.
\end{align}
In the continuum limit in the perturbative regime we
have~\cite{Jordan-Lee-Preskill_1111,Jordan-Lee-Preskill_1112}
\begin{equation} %onecolumn
m_0^2=\tilde{O}\Big(\frac{1}{a^{d-1}}\Big),\qquad
\lambda_0=\tilde{O}(1),\qquad
m_0^2<0,
\end{equation}
%\begin{align}
%m_0^2&=\tilde{O}\Big(\frac{1}{a^{d-1}}\Big),\\
%\lambda_0&=\tilde{O}(1),\\
%m_0^2&<0,
%\end{align} %twocolumn
so
\begin{equation}
\phi_{\text{cl}}=\tilde{O}\bigg(
\frac{\mathcal{V}^{1/4}}{a^{(d-1)/2}}\bigg).
\label{classical-phi}
\end{equation}
On the other hand,
\begin{equation}
\sigma_0=
\tilde{O}\bigg(\sqrt{\frac{Z}{a^{d-1}}}\bigg),
\end{equation}
and from comparison with~\eqref{classical-phi}, we obtain
\begin{equation}
\phi_{\text{cl}}=\tilde{O}\bigg(
\frac{\mathcal{V}^{1/4}}{Z^{1/2}}\sigma_0
\bigg).
\end{equation}
Provided $Z$ is nonvanishing, we see that this
result leads to a quadratic improvement in $\mathcal{V}$
with respect to the more
general bound~\eqref{new-bound}.

\bibliography{Bibliography_Matteo-Turco.bib}

\end{document}